\documentclass[useAMS,usenatbib,usegraphicx]{mn2e}

\def\Sobs{S_\rmn{obs}}
\def\FWHM{\rmn{FWHM}}
\def\figurewidth{84mm}
\def\d{\rmn{d}}

\begin{document}

\title[Strong lensing of SMGs]
{Strong lensing of submillimetre galaxies: A tracer of foreground structure?}

\author[G. Paciga, D. Scott and E.~L. Chapin]{Gregory Paciga, Douglas Scott
 and Edward~L. Chapin\\
  Department of Physics \& Astronomy, University of British Columbia,
  6224 Agricultural Road, Vancouver, B.C., V6T 1Z1, Canada}


\maketitle
\begin{abstract}
  The steep source counts and negative $K$-corrections of bright
  submillimetre galaxies (SMGs) suggest that a significant fraction of
  those observed at high flux densities may be gravitationally lensed,
  and that the lensing objects may often lie at redshifts above 1,
  where clusters of galaxies are difficult to detect through other
  means.  In this case follow-up of bright SMGs may be used to
  identify dense structures along the line of sight.  Here we
  investigate the probability for SMGs to experience strong lensing,
  using the latest $N$-body simulations and observed source flux and
  redshift distributions.  We find that almost all high redshift
  sources with a flux density above 100\,mJy will be lensed, if they
  are not relatively local galaxies.  We also give estimates of the
  fraction of sources experiencing strong lensing as a function of
  observed flux density.  This has implications for planning follow-up
  observations for bright SMGs discovered in future surveys with
  SCUBA-2 and other instruments.  The largest uncertainty in these
  calculations is the maximum allowed lensing amplification, which is
  dominated by the presently unknown spatial extent of SMGs.
\end{abstract}

\section{Introduction}
\label{sec:Introduction}

Sources that are picked up in deep extragalactic millimetre (mm) and
submillimetre (submm) wavelength surveys are far from being ordinary
galaxies. Due to strong evolution, and also driven by the typically
coarse angular resolution of current surveys at these wavelengths,
mm/submm sources are rare and luminous compared with well-studied
populations of optical galaxies. Submm galaxies (SMGs) are usually
interpreted as an early rapidly star-forming phase, perhaps driven by
a major merger, in the sequence that will ultimately become a massive
elliptical galaxy \citep[e.g.,][and references therein]{Blain02}.

However, gravitational lensing can also play a role, such that some
fraction of SMGs may be more typical, lower-luminosity galaxies at
high redshift ($z>1$) that are strongly amplified by lensing. This is
already a well known effect for quasars, where the steepening source
counts make lensing of increasing importance for the brightest objects
\citep[e.g.,][]{Kochanek04}.  In addition, the prevalence of giant
optical arcs is partly explained by the high redshifts of the sources
\citep{Wu06}. There are many anecdotal examples of SMGs that are
either known or strongly suspected to be lensed. Examples include
sources seen through targeted rich galaxy clusters, most spectacularly
in Abell 2218 \citep{Kneib96, Swinbank03}, and MS0451-03
\citep{Borys04}, as well as the brightest source in the central Hubble
Deep Field North, usually called HDF850.1 \citep{Dunlop04}, the
brightest source in the entire GOODS-North field, referred to as GN20
(Pope et al., submitted), and other bright well-studied sources, such
as SMM\,J14011+0252 \citep{Ivison01, Smail05}, MIPS\,J142824.0+352619
\citep{Borys06}, and SMM\,J02399-0136 \citep{Ivison98}.  It has long
been recognized \citep[e.g.,][]{Blain93} that the steep source counts
and high redshifts of SMGs make the brightest ones prime lensed
candidates.

Despite some efforts to model the lensing contribution to the source
counts \citep[e.g.,][]{Perrotta02, Perrotta03, Negrello07} the
predictions for SMG lensing are still very uncertain. The purpose of
the present paper is to improve this situation, using the latest
numerical models, and to try to quantify the uncertainties. In the
process of this investigation we will also be able to address a
related question -- namely whether bright SMGs could be used as
tracers of galaxy clusters at redshifts that are high enough to be
challenging for other techniques.

In order to carry out this study we will need several ingredients,
including the unlensed submm source counts and the redshift
distribution. However, we start in section~\ref{sec:ProbDensity} by
considering the probability distribution for lensing amplification
along random lines of sight. We then look at the observed and modelled
source counts for SMGs in section~\ref{sec:SourceCounts} and use these
to calculate the source counts after lensing in
section~\ref{sec:Calculating}, also exploring the dependence on the
redshift distribution of sources.  A major simplification of our
approach is to assume that the {\it shape\/} of the source counts is
independent of redshift; in order to test this we calculate in
section~\ref{sec:evolution} lensing expectations from a specific
evolutionary model that is consistent with a range of current IR and
sub-mm data.  Finally, we consider the uncertainties in the various
components of our analysis in section~\ref{sec:Uncertainties} and
interpret our results in the context of upcoming submm surveys in
section~\ref{sec:Predictions}.

\section{Probability Density Function}
\label{sec:ProbDensity}

\subsection{Millennium Simulation}
\label{sec:Millennium}

The first ingredient we will need for our study is the statistical
distribution of lensing amplifications. \citet{Wang02} presented a
`universal probability distribution function' valid for all
cosmologies and redshifts. However, it is only applicable to weak
lensing, with a maximum magnification much less than 10. As we are
interested in strong lensing effects leading to the very brightest
sources, where we will find a mean magnification far beyond the weak
lensing regime, this is unsuitable for our purposes. Other relevant
studies \citep{Perrotta02, Negrello07} have used largely analytic
approximations, have also focused on weak lensing \citep{Jain00, Keeton04},
or have relied on $N$-body simulations \citep{B98I, W92} which have
since been dwarfed in size by the Millennium Simulation
\citep{Springel05}, a computational run containing approximately
$10^{10}$ particles
of mass $8.6 \times 10^8 {\rm M}_{\odot}/h$ in a cube of side 500
Mpc/$h$.

\citet{Hilbert07} studied lensing amplification probability
distributions using a ray tracing approach with the Millennium
Simulation, and a $\Lambda$CDM cosmology with parameters
$\Omega_\rmn{M} = 0.25$, $\Omega_\Lambda = 0.75$, $h=0.73$, $n=1$, and
$\sigma_8=0.9$. They calculate optical depths $\tau^\rmn{I}_\mu$ and
$\tau^\rmn{S}_\mu$ for an image having been magnified by a factor
$\mu$.  The former gives the optical depth assuming a uniform
distribution of images, whereas the latter assumes a uniform
distribution of sources. Since we are lensing an approximately random
background of submm galaxies, henceforth we will use only
$\tau^\rmn{S}_\mu$ and its related probability density function (PDF)
$p^\rmn{S}(\mu)$, omitting the superscript `$\rmn{S}$' from now on.

Notice that this is not strictly the probability for a source undergoing a
total magnification $\mu$, but only for one image of a source.  There is
some ambiguity here over whether we should be considering the total
amplification for all images -- this will depend on how SMGs are
detected, the resolution of the imaging and on details of individual lensing
models.  Having pointed out this ambiguity, we will simply ignore it for the
rest of the paper.

Given $p(\mu)$, the probability that a source has been amplified by a
factor $\mu$ in the range ($\mu$, $\mu+\d\mu$) is then $p(\mu) \d\mu$.
Though the scalar magnification $\mu$ has a sign associated with it,
where negative values indicate a change in image polarity \citep[for a
review see, e.g.,][]{Schneider92, Courbin02}, the behaviour of the PDF
at $\mu < 1$ will allow us to take $\mu > 0$.  We will require that
the PDF be normalised:
\begin{equation}
\label{eqn:pdf-normalisation}
\int_{0}^{\infty} p(\mu) \d\mu = 1 .
\end{equation}
\noindent Since flux of a source must be conserved, sources amplified
in some directions must be deamplified in others. This flux
conservation constraint takes the form \citep{Hilbert07}
\begin{equation}
\label{eqn:flux conservation}
\int_{0}^{\infty} |\mu| p(\mu) \d\mu = 1 .
\end{equation}

The ray tracing techniques used essentially assume point sources,
which allows for arbitrarily high magnifications. The consideration of
extended objects imposes a maximum magnification limit, estimated by
\citet{Perrotta02} to be in the range 10--30 for 1--10~kpc sources at
redshifts 1--4. The applicability of this estimate to SMGs, however,
is uncertain, since the characteristic size of emission is
unknown. SMGs appear to be extended objects of a few kpc, but may well
be clumpy \citep{Pope05, Chapman03}, so small knots of bright
star-formation would have a higher potential for amplification by
lensing than the galaxy considered as a whole.  Perhaps the best empirical
constraint comes from
\citet{Younger07}, who recently constrained the sizes of the submm emitting
regions to $\la 10\,$kpc using interferometry.  Until we have much more
concrete information about the size distribution of SMGs there is little we can
do about the effects of finite source size. For now we continue
without a maximum magnification limit, but will consider in section
\ref{sec:UncLensing} how imposing such a constraint would
affect our results.

\subsection{Transform to other redshifts}
\label{sec:ProbTransform}

\citet{Hilbert07} give numerical data for a source plane at $z=2.1$,
as well as the peak position $\mu_\rmn{p}(z)$ and full width at half maximum
$\FWHM(z)$ of the distributions for other redshifts. We found that we
could transform the $z=2.1$ PDF to other redshifts using a
transformation of the form
\begin{eqnarray}
\label{eqn:transform}
\mu \longrightarrow & \mu^\prime & =
 \left( \frac{e^{b/\mu_\rmn{p}}-1}{e^{b/\mu}-1} \right)
 \mu_\rmn{p} + \delta \mu_\rmn{p}, \\
p(\mu) \longrightarrow & p^\prime(\mu) & = A p(\mu),
\end{eqnarray}
\noindent where $\mu_\rmn{p}$ is the peak amplification at the
reference $z$, $\delta \mu_\rmn{p}$ is the shift in the peak between
redshifts, $b$ is determined by requiring that we obtain the correct
FWHM, and the normalisation constant $A$ is
determined by requiring that we satisfy equation~(\ref{eqn:pdf-normalisation}).
The peak shifts to lower $\mu$ as $z$ increases, to
balance the high $\mu$ tail, allowing the PDF to satisfy
equation~(\ref{eqn:flux conservation}).

\begin{figure}
  \includegraphics[width=\figurewidth]{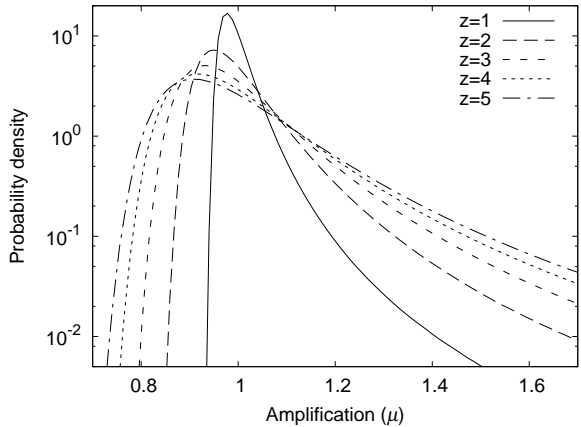}
  \caption{Family of lensing amplification PDF curves at different source
  redshifts, derived from \citet{Hilbert07}.}
  \label{fig:pdf-curves}
\end{figure}

This transformation preserves the asymptotic behaviour of the PDF, as
discussed in section~\ref{sec:ProbLimits}, which is most
crucial to our analysis.  We plot the resulting family of curves in
Fig.~\ref{fig:pdf-curves}.  For reference the values of the coefficients
are $(A,b)$ =
($0.473, -1.504$), ($0.968, -0.051$), ($1.383, 0.655$), ($1.681, 1.077$)
and ($1.889, 1.346$) for $z$ = 1, 2, 3, 4 and 5, respectively.

\subsection{Limiting behaviour}
\label{sec:ProbLimits}

The PDF for $z=2.1$ is seen to have a $\mu^{-3}$ tail, as predicted by
theory \citep{Schneider92} above an amplification of about
$\mu_\rmn{high} \simeq 20$. Thus, the data are fit to
\begin{equation}
\label{eqn:mu3}
p(\mu) = A_\rmn{high} \mu^{-3} \quad  \mbox{for $\mu > \mu_\rmn{high}$}  ,
\end{equation}
\noindent where $A_\rmn{high}$ is an adjustable parameter, and this
functional form can be used to extend the PDF to arbitrarily high
amplifications. The $\mu_\rmn{high}$ limit is adjusted for all
other redshifts according to the transformation described above.
Inaccuracy in the $\mu^{-3}$ behaviour above this limit may cause a
small kink in the function at the transition between the data and the
extrapolation, but we find this to be quite small. 

In the \citet{Hilbert07} data for $z=2.1$ there is a cut-off at about
$\mu=0.8$, where the PDF stops decreasing and reaches a near constant
plateau.  In this paper we are not interested in the deamplification
effects from the $\mu<1$ regime, so at most this low-$\mu$ behaviour
is important only for the normalisation of the PDF. We find that by
instead allowing the PDF to drop off like a Gaussian the normalisation
of the PDF is changed by about $10^{-6}$, and so is negligible for our
purposes.
We therefore assume that the PDF goes to zero like a Gaussian for
small $\mu$, i.e.,
\begin{equation}
\label{eqn:gaussian}
p(\mu) = A_\rmn{low} \exp \left( -\frac{(\mu-\mu_0)^2}{2 \sigma^2} \right)
 \quad \mbox{for $\mu < \mu_\rmn{low}$ },
\end{equation}
\noindent where $A_\rmn{low}$, $\mu_0$, and $\sigma$ are adjustable
parameters, and here $\mu_\rmn{low}$ is chosen to make as smooth a
transition from the data to the Gaussian as possible. The peak of the
distribution itself remains defined by the data. For example, for
$z=2.1$, the peak of the distribution is at $\mu \simeq 0.95$, while
the low Gaussian tail is not used until $\mu<\mu_\rmn{low}=0.8$.

This approach is in contrast to \citet{Perrotta03} who approximated the entire
weak lensing regime below $\mu \simeq 1.5$--2.0 as a Gaussian and the
entire strong lensing regime above the same limit by the $\mu^{-3}$
power-law.  Our approach is to use numerical data for a broad intermediate
range, giving a more accurate shape for the PDF, particularly for small
and intermediate amplifications.

\section{Source counts}
\label{sec:SourceCounts}

\subsection{Fits to observed counts}
\label{sec:SourceFits}

Functional fits to observed differential source counts of
submillimetre galaxies are given in \citet{Coppin06}, derived from the
SCUBA Half Degree Survey (SHADES) at $850\,\mu$m. They use both a
broken power-law of the form
\begin{equation}
\label{eqn:bpl}
\frac{\d N}{\d S} = \left\{ \begin{array}{ll} \frac{N^\prime}{S^\prime}
 \left( \frac{S}{S^\prime} \right) ^\alpha & \mbox{if $S<S^\prime$} , \\
\frac{N^\prime}{S^\prime} \left( \frac{S}{S^\prime} \right) ^\beta
 & \mbox{if $S>S^\prime$} , \end{array} \right.
\end{equation}
\noindent and a Schechter functional form
\begin{equation}
\label{eqn:schechter}
\frac{\d N}{\d S} = \frac{N^\prime}{S^\prime}
 \left( \frac{S}{S^\prime} \right) ^{\alpha+1} e^{-S/S^\prime}.
\end{equation}
\noindent with units of mJy$^{-1}$deg$^{-2}$. The differential
counts $\d N/\d S$ can be integrated from $S$ to infinity to obtain the
cumulative counts $N({>}\,S)$ with units of deg$^{-2}$. Currently there
are not enough high flux ($S>20$\,mJy) sources to prefer one function
over the other.

We have chosen to use the Schechter function as our primary estimate
of the source counts. The power-law drops off much slower than the
exponential part of the Schechter function, allowing for many more
intrinsically bright objects. As such, the most gradual power-law
consistent with the data can be used as an upper estimate to the
counts, which results in a far more conservative estimate of the
strong lensing probabilities. A lower limit to the source counts is
less helpful, as the corresponding uncertainty in the upper limit of
lensing probabilities is unknown.

The Schechter function parameters that we adopt are, in parallel with
\citet{Coppin06}, $N^\prime=1600$, $S^\prime=3.3$, and
$\alpha=-2$. For the maximal broken power-law we use $N^\prime=58$,
$S^\prime=9$, $\alpha=-2.2$, and $\beta=-4.2$.  The contribution to
the overall background (i.e.\ $\int S(dN/dS)dS$) from a power-law
number counts distribution diverges at the faint end if the slope
$\alpha\,{\le}\,-2$ (in equation~(\ref{eqn:bpl})).  To prevent
overproducing the submm background, we therefore set a limit on the
broken power-law of $S\ge0.2\,$mJy.

\subsection{Phenomenological models}
\label{sec:SourceModels}

\citet{Lagache03} used a phenomenological approach to model galaxy
evolution and predict number counts at a variety of wavelengths,
including $850\,\mu$m, and corrected this to include constraints from
\emph{Spitzer} data in \citeyear{Lagache04}. Their standard model is
roughly consistent with \citet{Coppin06} in the range where data from
SHADES exist, but decreases much more slowly above about 50\,mJy. This
is due entirely to local bright Euclidean counts, for which $\d N/\d S
\propto S^{-5/2}$. \citet{Chary01} similarly have a model for the
source counts that has been adjusted to include the \emph{Spitzer}
data. The results of their model at $850\,\mu$m is very similar to
\citet{Lagache04}, including the bright Euclidean part as well.

Normalising a Euclidean source count estimate to the SCUBA Local
Universe Galaxy Survey \citep[SLUGS;][]{Dunne00}, we find agreement
with the high flux behaviour in both the \citet{Lagache04} and
\citet{Chary01} models. Since the Euclidean part of the source counts
comes from low redshift sources, we do not expect there to be any
significant lensing here.  Subtracting the $S^{-5/2}$ part from either
of these models to estimate the high redshift population, we find that
the high flux fall-off is very similar to a Schechter function. This
is only a first-order approximation, as in principle the source counts
change for each redshift slice. We will investigate redshift
dependence in section \ref{sec:Comparing} and the change in the shape
of the counts with redshift in \ref{sec:UncCounts}.  We will also
compare with the results from an evolutionary model in
section~\ref{sec:evolution}.  But for now we will assume a single screen of
sources with counts independent of redshift.

\begin{figure}
  \includegraphics[width=\figurewidth]{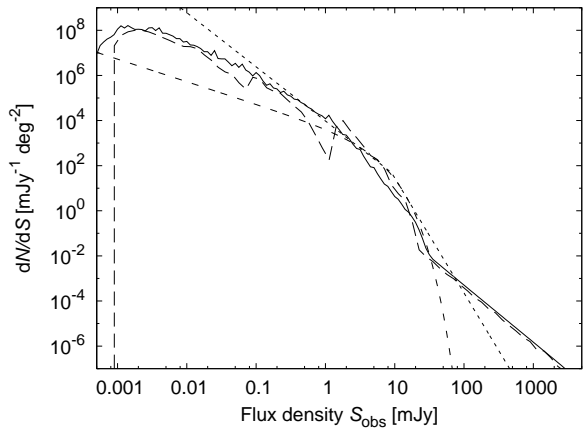}
  \caption{Differential number counts prior to applying lensing. Solid
    and long-dashed lines are the models of \citet{Lagache04} and
    \citet{Chary01}, respectively. Short-dashed and dotted lines are
    the Schechter functional form and broken power-laws from
    \citet{Coppin06}.}
  \label{fig:compare-counts}
\end{figure}

Between the low flux limit of the models at about $1\,\mu$Jy and the
knee at several mJy, both models predict intermediate counts between
the Schechter and broken power-law fits. Fig.~\ref{fig:compare-counts}
compares these functional fits to the models of \citet{Lagache04} and
\citet{Chary01}. This justifies our use of the Schechter and power-law
functional forms as our `best' and `maximal' estimates, respectively. The
lensed source count estimates in this paper at high flux densities
should be considered to be {\it in excess of\/} 
known local (e.g., \emph{IRAS}) galaxies, unless
otherwise stated. We will see that, as expected, highly amplified
sources make an increasing contribution at brighter fluxes.

\section{Calculating lensing effects}
\label{sec:Calculating}

\subsection{Observed source counts}
\label{sec:CalculatingCounts}

If an object with flux density $S$ is amplified through lensing by a
factor $\mu$, the object is observed with flux density $\Sobs = \mu S$
\citep{Schneider92}. For a given observed flux $\Sobs$, we run through
a range of magnifications, find the corresponding $S$, and assign
$p(\mu) \d N/\d S$ to be the contribution to the lensed differential
source counts from that $\mu$.  In this way we can express the
differential source counts after lensing as
\begin{equation}
\label{eqn:dNdSobs}
\frac{\d N}{\d\Sobs} \left( \Sobs \right) = \int_0^\infty \frac{\d N}{\d S}
 \left( \frac{\Sobs}{\mu} \right) p(\mu) \d\mu .
\end{equation}

\begin{figure}
  \includegraphics[width=\figurewidth]{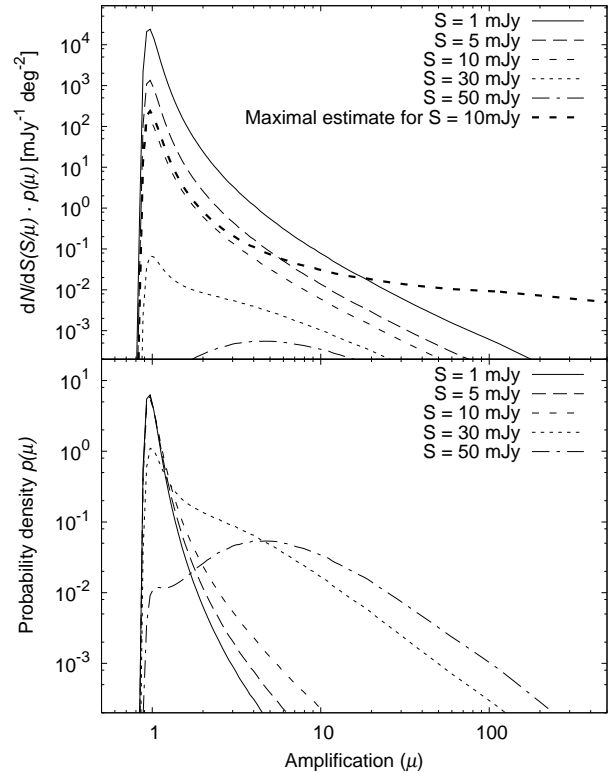}
  \caption{Magnification distributions. The contribution to $\d
  N/\d\Sobs$ made by each $\mu$ for various $\Sobs$ at $z=2$ (top) and
  normalised by the total $\d N/\d\Sobs$ (bottom).
  The bottom panel shows how the high $\mu$ tail dominates over the
  $\mu\simeq1$ peak for all $S\ga30\,$mJy.
  The additional
  thick line in the top panel is the `maximal' estimate for
  $\Sobs=10\,$mJy (the difference is typical for all $\Sobs$).}
  \label{fig:z2-amp-combo}
\end{figure}

For different values of $\Sobs$
we plot the contribution to the above integral as a
function of $\mu$ in Fig.~\ref{fig:z2-amp-combo}. We do this
specifically for sources at $z=2$, as well as showing a separate panel
with results normalised by the total contribution from all $\mu$. In the
normalised case, the line can be interpreted as a local $p(\mu)$ for
that $\Sobs$.  Although we have only plotted results for
$z=2$, other redshift choices look qualitatively similar.

\begin{figure}
  \includegraphics[width=\figurewidth]{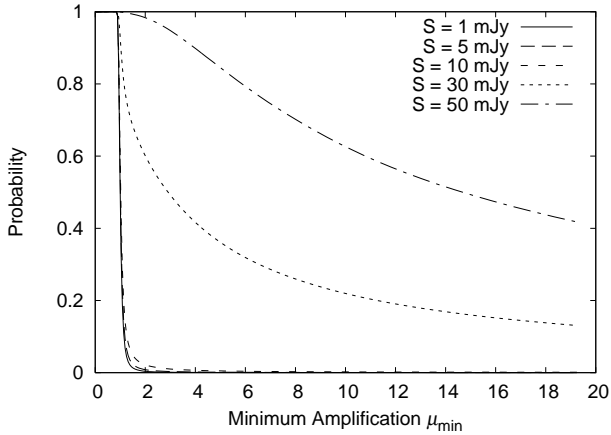}
  \caption{Combined contribution to $\d N/\d\Sobs$ from all
    magnifications above $\mu_\rmn{min}$ for sources at $z=2$.}
  \label{fig:z2-mmu}
\end{figure}

We can also examine the contribution to the source counts coming from
various magnifications by specifying a minimum $\mu$ that we might be
interested in.  The fractional contribution to $\d N/\d\Sobs$ from
$\mu>\mu_\rmn{min}$ is found simply by changing the lower limit on the
integral in equation~\ref{eqn:dNdSobs} from $0$ to $\mu_\rmn{min}$, and
normalising by the full $\d N/\d\Sobs$. This is plotted in
Fig.~\ref{fig:z2-mmu}.  It can be seen that at high flux densities a
large portion of sources will have significant amplifications.

\subsection{Comparing Redshifts}
\label{sec:Comparing}

\begin{figure}
  \includegraphics[width=\figurewidth]{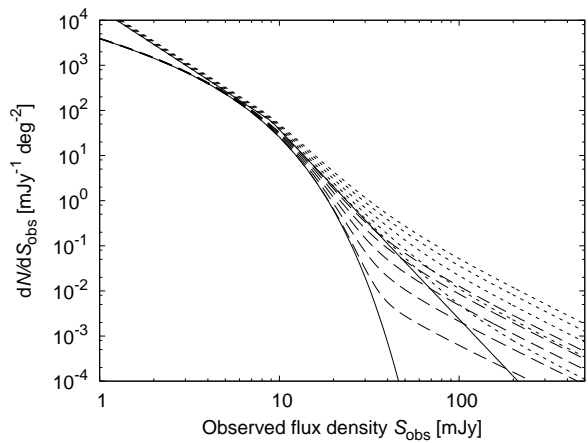}
  \caption{Lensed differential source counts found using equation
  \ref{eqn:dNdSobs}. Dashed lines represent counts derived from the best
  Schechter function (lower solid line) and dotted lines those from
  the maximal power-law (upper solid line).  Each set of curves is for
  sources placed at $z=1$, $2$, $3$, $4$, and $5$ from bottom to
  top.}
  \label{fig:num-combo}
\end{figure}

Lensing PDFs and a $\d N/\d S$ function can be combined at arbitrary
redshifts in the range $0.5 < z < 5.7$ (where \citet{Hilbert07}
quantify their lensing simulation).  The differential source counts of
the lensed populations are found via equation~\ref{eqn:dNdSobs} and
shown in Fig.~\ref{fig:num-combo} for a range of source redshifts. One
can see that the lensing tail is relatively important above about
20\,mJy and more so for higher redshift submm galaxies.

The cumulative source counts,
\begin{equation}
\label{eqn:cumulative}
N({>}\,S) = \int_S^\infty \frac{\d N}{\d\Sobs} \d\Sobs
\end{equation}
\noindent are used to determine the density of sources above a given
flux on the sky. This is a useful quantity for making observational
predictions for surveys.

\begin{figure}
  \includegraphics[width=\figurewidth]{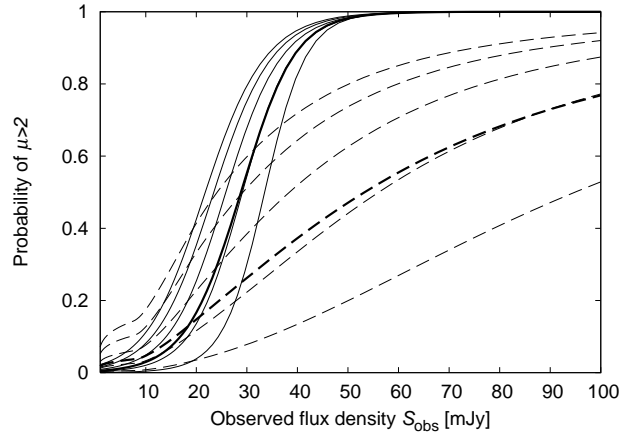}
  \caption{The fraction of $\d N/\d\Sobs$ due to magnifications
    greater than $2$, which can be interpreted as the probability of a
    lensing event as a function of $\Sobs$. The solid lines use the
    Schechter source counts and dashed lines use the maximal broken
    power-law, resulting in a minimal estimate of the lensing
    probability. We show (from bottom to top) $z=1$, $2$, $3$, $4$,
    and $5$ as thin lines and the average over all redshifts as a
    thick line in each case.}
  \label{fig:mmu-combo}
\end{figure}

Examining the contribution to source counts as a function of
$\mu_\rmn{min}$, as described above, we set, somewhat arbitrarily
(though consistent with \citealt{Negrello07} and \citealt{Perrotta02}), a
characteristic magnification for lensing events of
$\mu_\rmn{min}=2$. In Fig.~\ref{fig:mmu-combo} we plot the
contribution as a function of $\Sobs$ for various redshifts, as well
as the `best' and `maximal' estimates for the average over all
redshifts. This quantifies the probability that an object observed at
some $\Sobs$ has been lensed with a magnification greater than
$2$. One could easily obtain similar results for any other choice of
minimal amplification.

There have been several attempts to characterise the redshift
distribution of SMGs.  Though phenomenological models do exist
\citep[e.g.,][]{Granato01} we focus on estimates based purely on
observational data. \citet{Chapman05} obtained a large number of
spectroscopic redshifts for radio selected galaxies, while
\citet{Pope06} obtained both spectroscopic and photometric redshifts
in the GOODS-North survey, and \citet{Aretxaga07} used radio/submm
photometric redshifts in the SCUBA Half Degree Survey (SHADES). All
three are consistent with a Gaussian distribution centred at $z=2.2
\pm 0.1 $ and with $\sigma=0.8 \pm 0.1$. This is the distribution we use to
obtain the average over redshift in Fig.~\ref{fig:mmu-combo}. The
results are nearly identical when varying the peak redshift by
$\pm0.1$. Fig.~\ref{fig:mmu-combo} shows that for sources with $\Sobs
\simeq 20\,$mJy the probability of strong lensing is already 10--20
per cent for any reasonable redshift distribution.  Empirical evidence
suggests that this may be a conservative estimate, since at the moment
all sources detected with $\Sobs \ga 20\,$mJy at $850\,\mu$m have been
claimed to be lensed.

\begin{figure}
  \includegraphics[width=\figurewidth]{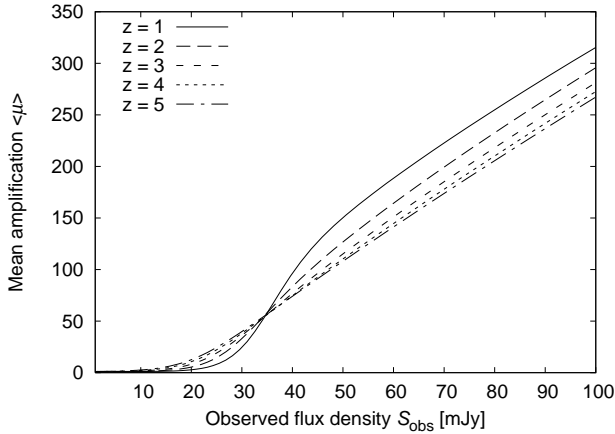}
  \caption{Mean magnification as a function of $\Sobs$.}
  \label{fig:ave}
\end{figure}

We can also look at the average amplification as a function of
observed flux density, given by
\begin{equation}
\label{average}
\langle\mu\rangle = \frac{ \int_0^\infty \mu \, \frac{\d N}{\d S}
 (\Sobs / \mu) \, p(\mu) \, \d\mu }
    { \int_0^\infty \frac{\d N}{\d S} (\Sobs / \mu) \, p(\mu) \, \d\mu } .
\end{equation}
\noindent
This is shown in Fig.~\ref{fig:ave} across a range of redshifts. One
can see a change in behaviour at higher and lower flux densities,
depending on the source redshift. Below about 35\,mJy the mean
amplification increases with redshift, while above $35\,$mJy it
decreases with redshift.  This is due to the change in dominance
between the $\mu \simeq 1$ peak and the high $\mu$ tail, an effect
that can be seen in the bottom panel of Fig.~\ref{fig:z2-amp-combo}.

\section{Uncertainties}
\label{sec:Uncertainties}

\subsection{Lensing probabilities}
\label{sec:UncLensing}

\citet{Hilbert07} discuss how they may have underestimated the lensing
properties by about 15 per cent, based on replacing the principle
lensing objects in their ray tracing by analytical haloes of the
\citet*{Navarro97} form.  This is surely a conservative estimate of
the effects since baryons are neglected.  Most large-scale
simulations, like the Millennium Simulation, include only
collisionless dark matter and not the much more complicated
dissipative baryons. Clumping due to baryons is bound to increase
strong lensing \citep[e.g.,][]{B98II} and so the normalisation of the
high $\mu$ tails that we use should be considered as a lower limit.
Recent studies \citep[e.g.,][]{Wambsganss08,Hilbert08} suggest that
structure due to baryon cooling and dissipation increase the incidence
of strong lensing by at least 25 per cent.

However, the effect of finite source size goes in the opposite
direction.  For extended sources there is an upper limit on the
magnification, as mentioned in section~\ref{sec:ProbDensity}. The
effect of this on the lensing probabilities can be significant,
depending on the counts and on details of the size distribution.  We
can use Fig.~\ref{fig:ave} to gauge how important this might be. If we
expect the limit to be ${\sim}\,30$, then the effect will be negligible
for sources with $\Sobs \la 30\,$mJy, but increasingly important for
brighter sources.

\begin{figure}
  \includegraphics[width=\figurewidth]{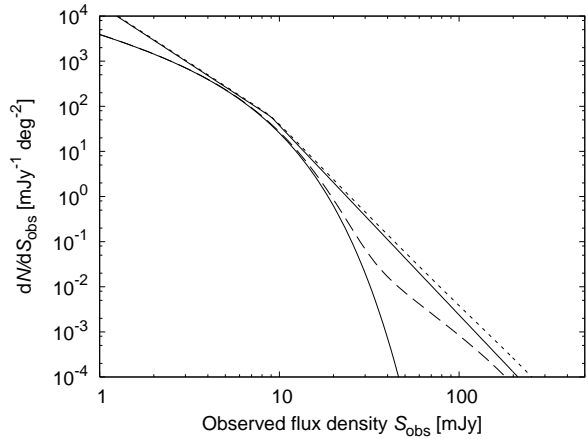}
  \caption{Differential source counts including the effects of lensing,
  {\it but\/} with a hard upper limit on the amplification of $\mu<30$.
  Here we have plotted lensed counts for the Schechter (dashed) and
  power-law (dotted) models, for the average over all source redshifts.
  This should be compared with Fig.~\ref{fig:num-combo}, making it clear
  how dramatic the effects are at the bright end.}
  \label{fig:counts30}
\end{figure}

\begin{figure}
  \includegraphics[width=\figurewidth]{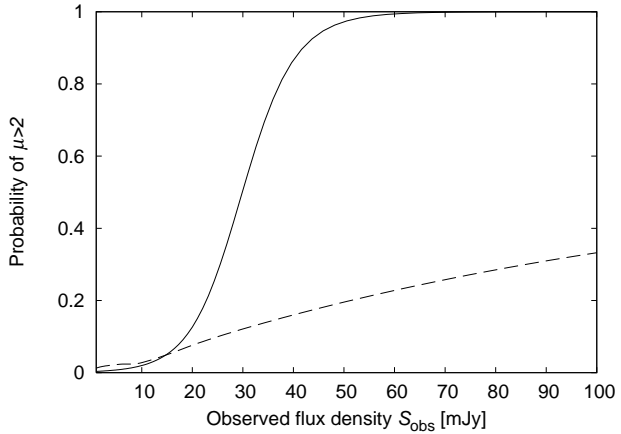}
  \caption{Fraction of the differential counts having amplification
  $\mu>2$, but restricted to lensing distributions with maximum
  of $\mu=30$.  Here we plot the results for the Schechter (solid) and
  power-law (dashed) counts models for the average over all source
  redshifts.  This plot should be compared with Fig.~\ref{fig:mmu-combo}.}
  \label{fig:prob30}
\end{figure}

We further show the importance of such a cut-off in Fig.~\ref{fig:prob30},
where we have calculated the differential source counts using the lensing
PDF, but not allowing any events at $\mu>30$.  One can see significant
changes for the brightest counts compared with Fig.~\ref{fig:num-combo}.
We also show the changes in the probability of strong lensing as a function
of observed flux density in Fig.~\ref{fig:mmu-combo}, again using a
maximal amplification of $\mu=30$.  One can see that the fraction of
strongly lensed sources is significantly reduced, particular for the
power-law counts model.

Is 30 a reasonable limit for lensing amplification?  One indication
that there is not a hard cut-off in $\mu$ is that the SMG lensed by
the cluster Abell~2218 has been estimated to have a (combined)
amplification of ${\simeq}\,45$ \citep{Kneib04}.  However, since there
is currently only vague size information for the submm-bright regions
of SMGs, there are no strong constraints on the maximum amplifications
that are possible, nor on the shape of any distribution at the high
$\mu$ end.  At this point we believe this should be regarded as an
empirical question that can only be answered by future high-resolution
studies with instruments such as ALMA.

Another potential source of systematic uncertainty lies in the fact that
there is only one Millennium Simulation, and it was not carried out for
the currently best-fitting cosmological model.  In particular the value
of $\sigma_8$ used (the normalization at a scale of $8\,h^{-1}$Mpc) was
0.9, while the {\it WMAP\/}5 value is more like 0.8 \citep{Dunkley08}.  
A smaller normalization can have a significant effect in reducing the
incidence of strong lensing \citep[e.g.,][]{Li07,Fedeli08}.  However, it is
difficult to assess precisely what affect this might have on our
estimates, for several reasons.  Firstly, the full simulations are not
available for other models.  Secondly, there are other differences in
the current best-fitting model, e.g.~$\Omega_{\rm M}$ and the primordial
fluctuation slope, $n$ (a red tilt means there are {\it lower\/}
fluctuations at galaxy and group scales than in the Millennium
simulation).  And finally, there is still some uncertainty
about the precise value of $\sigma_8$, with the statistics of giant
arcs suggesting that higher values may be a better fit
\citep[e.g.,][]{Li07,Fedeli08}.

\subsection{Source counts}
\label{sec:UncCounts}

As already discussed, we have taken the shallowest power-law consistent with
the SHADES data as found in \citet{Coppin06} and used that as an upper
estimate on the
source counts at the bright end. At high flux densities, this upper
estimate and our best guess Schechter function give significantly
different results, with the `maximal' counts providing the lower
estimate for lensing effects.

It is possible that the observed counts already include some
magnification bias. This being the case, our estimation of the
expected number of bright sources will be too high, since these
already amplified objects will be lensed further by our procedure in
Section~\ref{sec:Calculating}.  To attempt to compensate for this, we
looked at how the slope of the differential number counts changes when
lensing is applied {\it only\/} at the flux densities where the data
exist.  Using the Schechter function, we found a negligible change,
from which we conclude that lensing effects in this region are too
small to change our results. However, for the broken power-law, the
slope after the break ($\beta$ in equation~\ref{eqn:bpl}) is much more
sensitive to changes introduced by lensing. Above the knee,
\citet{Coppin06} found $\beta=5.1 \pm 0.9$.  Fitting
equation~\ref{eqn:bpl} to the lensed source counts for $\Sobs \la 25$,
where the counts still follow a broken power-law and SHADES data
exist, we find $\beta=4.9$. This is still well within the uncertainty
in the original broken power-law. In particular, the limit of
$N({>}\,22\,\rm{mJy})<17\,\rm{deg}^{-2}$ found by \citet{Coppin06} is
maintained by the maximal broken power-law even after lensing.

We have been making the simplifying assumption that the {\it shape\/}
of the submm counts is independent of redshift, even though the amount
of lensing might follow a particular source redshift distribution.
The main difficulty in carrying out estimates within more complicated
models is that there are only relatively weak constraints on how the
shape of the counts might vary with redshift.  The best available
evidence comes from a study of luminosity evolution among SMGs in the
GOODS-North field by \citet*{WPS07}, which builds
on earlier suggestions that the brightest SMGs tend to be at higher
redshift.  This appears to run counter to the idea of hierarchical
clustering, except of course it is most likely complications with the
baryons within galaxy formation that lead to this `down-sizing'.
\citet{WPS07} find evidence for 2 populations
separated in luminosity, each having a different luminosity function
slope and redshift dependence.  To the extent that the negative
$K$-correction makes up for distance dimming for SMGs, the submm flux
density is a proxy for luminosity, with the break occurring near
$5\,$mJy: fainter SMGs have a shallower slope and evolve more gently;
brighter SMGs have a steeper slope and evolve more strongly with
redshift.  Using this approximation we find that the phenomenological
models of \citet{WPS07} do indeed lead to SMG
counts which vary with redshift.  However, the results are (within the
fairly wide error bars) consistent with our `maximal' model.

At the moment the available SMG data involve modest sample sizes,
limited dynamic range and incomplete redshift information.  As more is
learned about how the counts evolve with redshift it will be possible to
carry out more accurate calculations of the lensing effects.

\section{Comparison with an evolutionary model}
\label{sec:evolution}

A short-coming of our analysis is the assumption of a fixed shape for
the source counts at all redshifts.  Naive expectations for galaxy
formation are that the most extreme SMGs would be rarer at high
redshift.  However, the idea of `downsizing' has been shown to apply
to SMGs (Wall, Pope \& Scott 2008), making the `hyper-LIRGs' more
prevalent at earlier times.  Hence, to properly assess these effects
requires a more detailed model which includes the evolution of SMGs.
We choose a particular set of simulations based on the local
far-infrared colour-luminosity distribution described in Chapin,
Hughes \& Aretxaga (2009) including evolution that reproduces the
observed 850\,\micron\ source counts and redshift distribution (see
the complete description in Appendix~\ref{model}). However, since the
existing empirical information is so meager, the model's behaviour at
high redshift can only be considered as indicative.  Moreover, since
the model is based on a particular simulation, then it is limited for
rare objects at the bright end.  Nevertheless, its use should enable us
to assess the effect on our estimates of allowing the number count
shape to vary as a function of redshift in a way which is consistent
with current data.

\begin{figure}
  \includegraphics[height=\figurewidth,angle=-90]{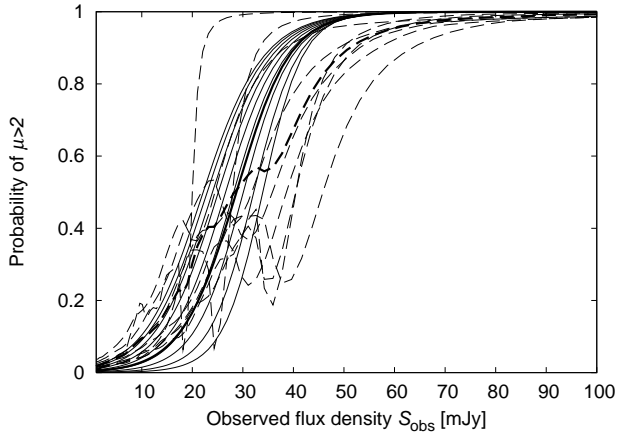}
  \caption{Dashed lines are the fraction of $\d N/\d\Sobs$ due to
    magnifications greater than $2$ using the model from
    Appendix~\ref{model} in redshift bins of $\Delta z = 0.5$ for $0.5
    < z < 5.0$. Solid lines are from the same redshift intervals using
    the Schechter function.  The thick lines are the average over all
    redshifts.}
  \label{fig:chapin}
\end{figure}

The simulation we utilise contains over 8 million sources between $z=0$ and 5.
We divided these sources into bins of width $\Delta z = 0.5$ and
determined the differential source counts $\rmn{d}N/\rmn{d}S$ for each
bin, to which we could apply the lensing probabilities in the same way
as before.  The probability of a lensing event as a function of
$\Sobs$ is shown in Fig.~\ref{fig:chapin}, and compared to the Schechter
function using the same redshift bins.

It can be noted that the evolutionary model produces considerably more
variation in lensing probability compared to the analytical Schechter
function.  Nonetheless, the probability is still greater than about
0.3 for any object with an observed flux greater than about 20\,mJy,
which is comparable to the results we found for a single counts shape.
For dimmer objects, where we may need to be conservative about what
sources deserve closer attention, the chances of a lensing event are
actually slightly better with the evolutionary model.  So the
conclusion is that although our analytical approximation misses some
of the structure in the redshift distribution of SMGs, the end results
are quite similar.  Still, this should be considered as yet another
potential source of uncertainty in the lensing predictions, and
further motivation for gathering data which can distinguish among
details of the models.

\section{Discussion and Conclusions}
\label{sec:Predictions}

We can use our study to make approximate predictions for upcoming
submm surveys.  There are several relevant instruments, but we focus
on the surveys which are planned with the SCUBA-2 instrument
\citep{Holland06} on the James Clerk Maxwell Telescope.  There are 2
relevant $850\,\mu$m programmes: the SCUBA-2 Cosmology Legacy Survey,
S2CLS; and the SCUBA-2 `All Sky' Survey, SASSy \citep{Thompson07}.

We have also carried out estimates for surveys at shorter wavelengths,
of the sort which might be performed with the $450\,\mu$m array of
SCUBA-2, or the SPIRE instrument on the {\it Herschel} satellite (operating
at $250$, $350$ and $500\,\mu$m).  While
it is clear that there may be many examples of strong lenses in such wide
surveys, the fraction of bright sources which are lensed is significantly
lower than at longer wavelengths.  If one wants to find such `monsters',
either as probes of line-of-sight structure or for their own intrinsic
value, then one should turn to ground-based surveys in the $850\,\mu$m or
${\sim}\,1\,$mm windows.

The S2CLS plans to map approximately $20\,{\rm deg}^2$ to an RMS of
$0.7\,$mJy.  From our number counts model we estimate that there will
be about 96, 44 and 27 sources detected with $S\,{>}\,20$, 25 and
$30\,$mJy, respectively.  The total number of sources above these flux
limits which have $\mu\,{>}\,2$ will be about 15 (with most of them at the
high flux end) using our best
estimate from the Schechter function number counts. The numbers change only
slightly over this flux density range,
due to the probability of lensing increasing faster than
the decrease in source counts. Above $35\,$mJy the number of expected
lensed sources declines steadily. The lensed fractions are
considerably smaller using our `maximal' counts model, as can be seen
by referring to Fig.~\ref{fig:mmu-combo}.

There will of course always be bright sources which have negligible lensing
-- hence one would like to know how bright to go before the probability of
strong lensing is significant.  This can be determined using
Fig.~\ref{fig:mmu-combo}.
If one is prepared to accept a 1 in 3 chance of a source being
strongly lensed, then one should select sources observed with
$S\,{\ga}\,25\,$mJy.  If one would like the chance to be 1 in 2, then
that flux rises to about $30\,$mJy.

What would one do with such sources in practice?  Since strong lensing is
likely to come from either a galaxy cluster, or a massive galaxy, then
the existence of strong lensing implies strongly clustered structure
along that line of sight.  Hence for each sufficiently bright source
one would use follow-up observations
at other wavelengths to try to establish whether strong lensing was
likely, and then to see if one could find the structure which was
responsible for the lensing.  Multiple images or distorted
morphologies of optical counterparts would be ways of determining that
lensing was taking place -- these would naturally show up as part of
the procedure for trying to determine counterparts in deep data at
other wavelengths.  For SMGs where lensing was strongly suspected, one
would target the area to search for the presence of structure along
the line of sight, either with X-ray, Sunyaev-Zel'dovich or `red
cluster sequence' observations.  The advantage of this approach is
that the high redshift of the SMG sources means that it should be
feasible to find cluster (or proto-cluster) lenses at higher redshifts
than are easy to achieve with most other techniques.  Of course, the
selection effects for clusters found in this way may be complicated to
quantify.  Nevertheless, building up samples of $z\,{>}\,1$ clusters is
sufficiently important for understanding structure formation (as well
as constraining dark energy, etc.), that it is worth using every
available method.

SASSy is designed to make a shallow $850\,\mu$m map over
approximately $4{,}000\,{\rm deg}^2$, or one tenth of the sky, with
the possibility of extension to a larger area later.  The RMS of the
maps is planned to be around $30\,$mJy, so that a robust $5\sigma$
catalogue will have a limit around $150\,$mJy.  It may also be
possible to reduce this to nearer to $100\,$mJy using targetted repeat
observations for peaks in the maps.

The procedure for identifying strongly lensed sources in SASSy may be
a little different than for S2CLS, and because of the brighter flux
densities, the level of uncertainty in predictions of the number of
lensed sources will be considerably higher.  These predictions depend
strongly on the counts model, the normalization of the high
amplification tail of the lensing PDF, as well as the amplification cut-off
imposed by the finite source size for SMGs.  Since there are still
huge uncertainties in all of these factors, the expectations for SASSy
cover a wide range of possibilities.

Using the optimistic limit of $100\,$mJy for SASSy, the unlensed
model counts give approximately 1200 sources for the SASSy catalogue.
Many of these will be in the `Euclidean counts' regime, and hence
should be relatively easy to eliminate as lensing candidates.  These will often
be already well-known galaxies, with others typically being
in the {\sl IRAS\/}, {\sl Akari\/} or radio catalogues.  Colours
(e.g.,~$850\,\mu$m to radio) can be used to distinguish objects which
are likely to be at higher redshift and hence have a higher likelihood of
being lensed; such methods will also be necessary to
eliminate Galactic clouds.  Our best lensing estimate yields a total
of 1000 sources in SASSy with lensing amplification $\mu\,{>}\,2$, with
most of them being much more strongly lensed than this limit.  We expect
that these extremely lensed sources will be fairly easy to
distinguish from relatively nearby, intrinsically bright galaxies, and
hence the chances of there being structure along the line of sight to
such candidates will be very high.  Follow-up of these SMGs at other
wavelengths will also be easy, since they should be at least an order
of magnitude brighter than the typical SCUBA
sources which have been followed up in the past.

\section{Acknowledgments}
\label{Acknowledgments}

We would like to thank Alexandra Pope for many helpful discussions
during the course of this work.  Neal Dalal and Mattia Negrello
provided useful feedback on an early draft of this paper.  We also
thank Ranga-Ram Chary for providing his submillimetre galaxy models to
us. The models of \citeauthor{Lagache04} are freely available
online\footnote{http://www.ias.u-psud.fr/irgalaxies/Model/}. This work
was supported by the Natural Sciences and Engineering Research Council
of Canada.

\bibliography{ref-mnras}

\begin{thebibliography}{}

\bibitem[\protect\citeauthoryear{{Aretxaga}, {Hughes}, {Coppin}, {Mortier},
  {Wagg}, {Dunlop}, {Chapin}, {Eales}, {Gaztanaga}, {Halpern}, {Ivison}, {van
  Kampen}, {Scott}, {Serjeant} \& {Smail}}{{Aretxaga}
  et~al.}{2007}]{Aretxaga07}
{Aretxaga} I.,  {Hughes} D.~H.,  {Coppin} K.,  {Mortier} A.~M.~J.,  {Wagg} J.,
  {Dunlop} J.~S.,  {Chapin} E.~L.,  {Eales} S.~A.,  {Gaztanaga} E.,  {Halpern}
  M.,  {Ivison} R.~J.,  {van Kampen} E.,  {Scott} D.,  {Serjeant} S.,
  {Smail} I.,  2007, ArXiv Astrophysics e-prints

\bibitem[\protect\citeauthoryear{{Bartelmann}, {Steinmetz} \&
  {Weiss}}{{Bartelmann} et~al.}{1995}]{B98II}
{Bartelmann} M.,  {Steinmetz} M.,    {Weiss} A.,  1995, A\&A, 297, 1

\bibitem[\protect\citeauthoryear{{Bartelmann} \& {Weiss}}{{Bartelmann} \&
  {Weiss}}{1994}]{B98I}
{Bartelmann} M.,  {Weiss} A.,  1994, A\&A, 287, 1

\bibitem[\protect\citeauthoryear{{Blain} \& {Longair}}{{Blain} \&
  {Longair}}{1993}]{Blain93}
{Blain} A.~W.,  {Longair} M.~S.,  1993, \mnras, 264, 509

\bibitem[\protect\citeauthoryear{{Blain}, {Smail}, {Ivison}, {Kneib} \&
  {Frayer}}{{Blain} et~al.}{2002}]{Blain02}
{Blain} A.~W.,  {Smail} I.,  {Ivison} R.~J.,  {Kneib} J.-P.,    {Frayer} D.~T.,
   2002, Phys.~Rep., 369, 111

\bibitem[\protect\citeauthoryear{{Borys}, {Blain}, {Dey}, {Le Floc'h},
  {Jannuzi}, {Barnard}, {Bian}, {Brodwin}, {Men{\'e}ndez-Delmestre},
  {Thompson}, {Brand}, {Brown}, {Dowell}, {Eisenhardt} \& {Farrah}}{{Borys}
  et~al.}{2006}]{Borys06}
{Borys} C.,  {Blain} A.~W.,  {Dey} A.,  {Le Floc'h} E.,  {Jannuzi} B.~T.,
  {Barnard} V.,  {Bian} C.,  {Brodwin} M.,  {Men{\'e}ndez-Delmestre} K.,
  {Thompson} D.,  {Brand} K.,  {Brown} M.~J.~I.,  {Dowell} C.~D.,  {Eisenhardt}
  P.,    {Farrah} D.,  2006, \apj, 636, 134

\bibitem[\protect\citeauthoryear{{Borys}, {Chapman}, {Donahue}, {Fahlman},
  {Halpern}, {Kneib}, {Newbury}, {Scott} \& {Smith}}{{Borys}
  et~al.}{2004}]{Borys04}
{Borys} C.,  {Chapman} S.,  {Donahue} M.,  {Fahlman} G.,  {Halpern} M.,
  {Kneib} J.-P.,  {Newbury} P.,  {Scott} D.,    {Smith} G.~P.,  2004, \mnras,
  352, 759

\bibitem[\protect\citeauthoryear{{Chapin}, {Hughes}, } \&
  {Aretxaga}}{{Chapin} et~al.}{2009}]{Chapin09}
{Chapin} E.~L.,  {Hughes} D.~H.,  {Aretxaga} I., 2009, \mnras, in press

\bibitem[\protect\citeauthoryear{{Chapman}, {Blain}, {Ivison} \&
  {Smail}}{{Chapman} et~al.}{2003}]{Chapman03}
{Chapman} S.~C.,  {Blain} A.~W.,  {Ivison} R.~J.,    {Smail} I.~R.,  2003,
  \nat, 422, 695

\bibitem[\protect\citeauthoryear{{Chapman}, {Blain}, {Smail} \&
  {Ivison}}{{Chapman} et~al.}{2005}]{Chapman05}
{Chapman} S.~C.,  {Blain} A.~W.,  {Smail} I.,    {Ivison} R.~J.,  2005, \apj,
  622, 772

\bibitem[\protect\citeauthoryear{{Chary} \& {Elbaz}}{{Chary} \&
  {Elbaz}}{2001}]{Chary01}
{Chary} R.,  {Elbaz} D.,  2001, \apj, 556, 562

\bibitem[\protect\citeauthoryear{{Coppin}, {Chapin}, {Mortier}, {Scott},
  {Borys}, {Dunlop}, {Halpern}, {Hughes}, {Pope}, {Scott}, {Serjeant}, {Wagg},
  {Alexander}, {Almaini} \& {Aretxaga}}{{Coppin} et~al.}{2006}]{Coppin06}
{Coppin} K.,  {Chapin} E.~L.,  {Mortier} A.~M.~J.,  {Scott} S.~E.,  {Borys} C.,
   {Dunlop} J.~S.,  {Halpern} M.,  {Hughes} D.~H.,  {Pope} A.,  {Scott} D.,
  {Serjeant} S.,  {Wagg} J.,  {Alexander} D.~M.,  {Almaini} O.,    {Aretxaga}
  I.,  2006, \mnras, 372, 1621

\bibitem[\protect\citeauthoryear{{Courbin}, {Saha} \& {Schechter}}{{Courbin}
  et~al.}{2002}]{Courbin02}
{Courbin} F.,  {Saha} P.,    {Schechter} P.~L.,  2002, in {Courbin} F.,
  {Minniti} D.,  eds, Gravitational Lensing: An Astrophysical Tool Vol.~608 of
  Lecture Notes in Physics, Berlin Springer Verlag, {Quasar Lensing}.
pp~1--+

\bibitem[\protect\citeauthoryear{{Dunlop}, {McLure}, {Yamada}, {Kajisawa},
  {Peacock}, {Mann}, {Hughes}, {Aretxaga}, {Muxlow}, {Richards}, {Dickinson},
  {Ivison}, {Smith}, {Smail}, {Serjeant}, {Almaini} \& {Lawrence}}{{Dunlop}
  et~al.}{2004}]{Dunlop04}
{Dunlop} J.~S.,  {McLure} R.~J.,  {Yamada} T.,  {Kajisawa} M.,  {Peacock}
  J.~A.,  {Mann} R.~G.,  {Hughes} D.~H.,  {Aretxaga} I.,  {Muxlow} T.~W.~B.,
  {Richards} A.~M.~S.,  {Dickinson} M.,  {Ivison} R.~J.,  {Smith} G.~P.,
  {Smail} I.,  {Serjeant} S.,  {Almaini} O.,    {Lawrence} A.,  2004, \mnras,
  350, 769

\bibitem[\protect\citeauthoryear{{Dunne}, {Eales}, {Edmunds}, {Ivison},
  {Alexander} \& {Clements}}{{Dunne} et~al.}{2000}]{Dunne00}
{Dunne} L.,  {Eales} S.,  {Edmunds} M.,  {Ivison} R.,  {Alexander} P.,
  {Clements} D.~L.,  2000, \mnras, 315, 115

\bibitem[\protect\citeauthoryear{{Granato}, {Silva}, {Monaco}, {Panuzzo},
  {Salucci}, {De Zotti} \& {Danese}}{{Granato} et~al.}{2001}]{Granato01}
{Granato} G.~L.,  {Silva} L.,  {Monaco} P.,  {Panuzzo} P.,  {Salucci} P.,  {De
  Zotti} G.,    {Danese} L.,  2001, \mnras, 324, 757

\bibitem[\protect\citeauthoryear{{Haarsma} \& {Partridge}}{{Haarsma} \&
  {Partridge}}{1998}]{Haarsma98}
{Haarsma} D.~B.,  {Partridge} R.~B.,  1998, \apjl, 503, L5+

\bibitem[\protect\citeauthoryear{{Hilbert}, {White}, {Hartlap} \&
  {Schneider}}{{Hilbert} et~al.}{2007}]{Hilbert07}
{Hilbert} S.,  {White} S.~D.~M.,  {Hartlap} J.,    {Schneider} P.,  2007, ArXiv
  Astrophysics e-prints

\bibitem[\protect\citeauthoryear{{Holland}, {MacIntosh}, {Fairley}, {Kelly},
  {Montgomery}, {Gostick}, {Atad-Ettedgui}, {Ellis}, {Robson} \&
  {Hollister}}{{Holland} et~al.}{2006}]{Holland06}
{Holland} W.,  {MacIntosh} M.,  {Fairley} A.,  {Kelly} D.,  {Montgomery} D.,
  {Gostick} D.,  {Atad-Ettedgui} E.,  {Ellis} M.,  {Robson} I.,    {Hollister}
  M.,  2006, in Millimeter and Submillimeter Detectors and Instrumentation for
  Astronomy III. Edited by Zmuidzinas, Jonas; Holland, Wayne S.; Withington,
  Stafford; Duncan, William D.. Proceedings of the SPIE, Volume 6275, pp.
  62751E (2006). Vol.~6275 of Presented at the Society of Photo-Optical
  Instrumentation Engineers (SPIE) Conference, {SCUBA-2: a 10,000-pixel
  submillimeter camera for the James Clerk Maxwell Telescope}

\bibitem[\protect\citeauthoryear{{Ivison}, {Smail}, {Frayer}, {Kneib} \&
  {Blain}}{{Ivison} et~al.}{2001}]{Ivison01}
{Ivison} R.~J.,  {Smail} I.,  {Frayer} D.~T.,  {Kneib} J.-P.,    {Blain} A.~W.,
   2001, \apjl, 561, L45

\bibitem[\protect\citeauthoryear{{Ivison}, {Smail}, {Le Borgne}, {Blain},
  {Kneib}, {Bezecourt}, {Kerr} \& {Davies}}{{Ivison} et~al.}{1998}]{Ivison98}
{Ivison} R.~J.,  {Smail} I.,  {Le Borgne} J.-F.,  {Blain} A.~W.,  {Kneib}
  J.-P.,  {Bezecourt} J.,  {Kerr} T.~H.,    {Davies} J.~K.,  1998, \mnras, 298,
  583

\bibitem[\protect\citeauthoryear{{Jain}, {Seljak} \& {White}}{{Jain}
  et~al.}{2000}]{Jain00}
{Jain} B.,  {Seljak} U.,    {White} S.,  2000, \apj, 530, 547

\bibitem[\protect\citeauthoryear{{Keeton}, {Kuhlen} \& {Haiman}}{{Keeton}
  et~al.}{2005}]{Keeton04}
{Keeton} C.~R.,  {Kuhlen} M.,    {Haiman} Z.,  2005, \apj, 621, 559

\bibitem[\protect\citeauthoryear{{Kneib}, {Ellis}, {Smail}, {Couch} \&
  {Sharples}}{{Kneib} et~al.}{1996}]{Kneib96}
{Kneib} J.-P.,  {Ellis} R.~S.,  {Smail} I.,  {Couch} W.~J.,    {Sharples}
  R.~M.,  1996, \apj, 471, 643

\bibitem[\protect\citeauthoryear{{Kneib}, {van der Werf}, {Kraiberg Knudsen},
 {Smail}, {Blain}, {Frayer}, {Barnard} \&
  {Ivison}}{{Kneib} et~al.}{2004}]{Kneib04}
{Kneib} J.-P., {van der Werf} P.~P., {Kraiberg Knudsen} K.,
 {Smail} I., {Blain} A., {Frayer} D., {Barnard} V., {Ivison} R.,
  2004, \mnras, 349, 1211

\bibitem[\protect\citeauthoryear{{Kochanek}}{{Kochanek}}{2004}]{Kochanek04}
{Kochanek} C.~S.,  2004, ArXiv Astrophysics e-prints

\bibitem[\protect\citeauthoryear{{Lagache}, {Dole} \& {Puget}}{{Lagache}
  et~al.}{2003}]{Lagache03}
{Lagache} G.,  {Dole} H.,    {Puget} J.-L.,  2003, \mnras, 338, 555

\bibitem[\protect\citeauthoryear{{Lagache}, {Dole}, {Puget},
  {P{\'e}rez-Gonz{\'a}lez}, {Le Floc'h}, {Rieke}, {Papovich}, {Egami},
  {Alonso-Herrero}, {Engelbracht}, {Gordon}, {Misselt} \& {Morrison}}{{Lagache}
  et~al.}{2004}]{Lagache04}
{Lagache} G.,  {Dole} H.,  {Puget} J.-L.,  {P{\'e}rez-Gonz{\'a}lez} P.~G.,  {Le
  Floc'h} E.,  {Rieke} G.~H.,  {Papovich} C.,  {Egami} E.,  {Alonso-Herrero}
  A.,  {Engelbracht} C.~W.,  {Gordon} K.~D.,  {Misselt} K.~A.,    {Morrison}
  J.~E.,  2004, \apjs, 154, 112

\bibitem[\protect\citeauthoryear{{Navarro}, {Frenk} \& {White}}{{Navarro}
  et~al.}{1997}]{Navarro97}
{Navarro} J.~F.,  {Frenk} C.~S.,    {White} S.~D.~M.,  1997, \apj, 490, 493

\bibitem[\protect\citeauthoryear{{Negrello}, {Perrotta}, {Gonz{\'a}lez},
  {Silva}, {de Zotti}, {Granato}, {Baccigalupi} \& {Danese}}{{Negrello}
  et~al.}{2007}]{Negrello07}
{Negrello} M.,  {Perrotta} F.,  {Gonz{\'a}lez} J.~G.-N.,  {Silva} L.,  {de
  Zotti} G.,  {Granato} G.~L.,  {Baccigalupi} C.,    {Danese} L.,  2007,
  \mnras, 377, 1557

\bibitem[\protect\citeauthoryear{{Perrotta}, {Baccigalupi}, {Bartelmann}, {De
  Zotti} \& {Granato}}{{Perrotta} et~al.}{2002}]{Perrotta02}
{Perrotta} F.,  {Baccigalupi} C.,  {Bartelmann} M.,  {De Zotti} G.,
  {Granato} G.~L.,  2002, \mnras, 329, 445

\bibitem[\protect\citeauthoryear{{Perrotta}, {Magliocchetti}, {Baccigalupi},
  {Bartelmann}, {De Zotti}, {Granato}, {Silva} \& {Danese}}{{Perrotta}
  et~al.}{2003}]{Perrotta03}
{Perrotta} F.,  {Magliocchetti} M.,  {Baccigalupi} C.,  {Bartelmann} M.,  {De
  Zotti} G.,  {Granato} G.~L.,  {Silva} L.,    {Danese} L.,  2003, \mnras, 338,
  623

\bibitem[\protect\citeauthoryear{{Pope}, {Borys}, {Scott}, {Conselice},
  {Dickinson} \& {Mobasher}}{{Pope} et~al.}{2005}]{Pope05}
{Pope} A.,  {Borys} C.,  {Scott} D.,  {Conselice} C.,  {Dickinson} M.,
  {Mobasher} B.,  2005, \mnras, 358, 149

\bibitem[\protect\citeauthoryear{{Pope}, {Scott}, {Dickinson}, {Chary},
  {Morrison}, {Borys}, {Sajina}, {Alexander}, {Daddi}, {Frayer}, {MacDonald} \&
  {Stern}}{{Pope} et~al.}{2006}]{Pope06}
{Pope} A.,  {Scott} D.,  {Dickinson} M.,  {Chary} R.-R.,  {Morrison} G.,
  {Borys} C.,  {Sajina} A.,  {Alexander} D.~M.,  {Daddi} E.,  {Frayer} D.,
  {MacDonald} E.,    {Stern} D.,  2006, \mnras, 370, 1185

\bibitem[\protect\citeauthoryear{{Schneider}, {Ehlers} \& {Falco}}{{Schneider}
  et~al.}{1992}]{Schneider92}
{Schneider} P.,  {Ehlers} J.,    {Falco} E.~E.,  1992, {Gravitational Lenses}.
Springer-Verlag

\bibitem[\protect\citeauthoryear{{Smail}, {Smith} \& {Ivison}}{{Smail}
  et~al.}{2005}]{Smail05}
{Smail} I.,  {Smith} G.~P.,    {Ivison} R.~J.,  2005, \apj, 631, 121

\bibitem[\protect\citeauthoryear{{Springel}, {White}, {Jenkins}, {Frenk},
  {Yoshida}, {Gao}, {Navarro}, {Thacker}, {Croton}, {Helly}, {Peacock}, {Cole},
  {Thomas}, {Couchman}, {Evrard}, {Colberg} \& {Pearce}}{{Springel}
  et~al.}{2005}]{Springel05}
{Springel} V.,  {White} S.~D.~M.,  {Jenkins} A.,  {Frenk} C.~S.,  {Yoshida} N.,
   {Gao} L.,  {Navarro} J.,  {Thacker} R.,  {Croton} D.,  {Helly} J.,
  {Peacock} J.~A.,  {Cole} S.,  {Thomas} P.,  {Couchman} H.,  {Evrard} A.,
  {Colberg} J.,    {Pearce} F.,  2005, Nature, 435, 629

\bibitem[\protect\citeauthoryear{{Swinbank}, {Smith}, {Bower}, {Bunker},
  {Smail}, {Ellis}, {Smith}, {Kneib}, {Sullivan} \&
  {Allington-Smith}}{{Swinbank} et~al.}{2003}]{Swinbank03}
{Swinbank} A.~M.,  {Smith} J.,  {Bower} R.~G.,  {Bunker} A.,  {Smail} I.,
  {Ellis} R.~S.,  {Smith} G.~P.,  {Kneib} J.-P.,  {Sullivan} M.,
  {Allington-Smith} J.,  2003, \apj, 598, 162

\bibitem[\protect\citeauthoryear{{Thompson}, {Serjeant}, {Jenness}, {Scott},
  {Ashdown}, {Brunt}, {Butner}, {Chapin}, {Chrysostomou}, {Clark}, {Clements},
  {Collett}, {Coppin}, {Coulson} \& {Dent}}{{Thompson}
  et~al.}{2007}]{Thompson07}
{Thompson} M.~A.,  {Serjeant} S.,  {Jenness} T.,  {Scott} D.,  {Ashdown} M.,
  {Brunt} C.,  {Butner} H.,  {Chapin} E.,  {Chrysostomou} A.~C.,  {Clark}
  J.~S.,  {Clements} D.,  {Collett} J.~L.,  {Coppin} K.,  {Coulson} I.~M.,
  {Dent} W.~R.~F.,  2007, ArXiv e-prints, 704

\bibitem[\protect\citeauthoryear{{Wambsganss}}{{Wambsganss}}{1992}]{W92}
{Wambsganss} J.,  1992, \apj, 386, 19

\bibitem[\protect\citeauthoryear{{Wang}, {Holz} \& {Munshi}}{{Wang}
  et~al.}{2002}]{Wang02}
{Wang} Y.,  {Holz} D.~E.,    {Munshi} D.,  2002, \apjl, 572, L15

\bibitem[\protect\citeauthoryear{{Wu} \& {Chiueh}}{{Wu} \&
  {Chiueh}}{2006}]{Wu06}
{Wu} J.~M.,  {Chiueh} T.,  2006, \apj, 639, 695

\bibitem[\protect\citeauthoryear{{Younger}, {Fazio}, {Huang}, {Yun}, {Wilson},
  {Ashby}, {Gurwell}, {Lai}, {Peck}, {Petitpas}, {Wilner}, {Iono}, {Kohno},
  {Kawabe}, {Hughes} \& {Aretxaga}}{{Younger} et~al.}{2007}]{Younger07}
{Younger} J.~D.,  {Fazio} G.~G.,  {Huang} J.-S.,  {Yun} M.~S.,  {Wilson} G.~W.,
   {Ashby} M.~L.~N.,  {Gurwell} M.~A.,  {Lai} K.,  {Peck} A.~B.,  {Petitpas}
  G.~R.,  {Wilner} D.~J.,  {Iono} D.,  {Kohno} K.,  {Kawabe} R.,  {Hughes}
  D.~H.,    {Aretxaga} I.,  2007, ArXiv e-prints, 708

\end{thebibliography}


\begin{thebibliography}{}

\bibitem[\protect\citeauthoryear{{Aretxaga}, {Hughes}, {Coppin}, {Mortier},
  {Wagg}, {Dunlop}, {Chapin}, {Eales}, {Gazta{\~n}aga}, {Halpern}, {Ivison},
  {van Kampen}, {Scott}, {Serjeant} \& {Smail}}{{Aretxaga}
  et~al.}{2007}]{Aretxaga07}
{Aretxaga} I.,  {Hughes} D.~H.,  {Coppin} K.,  {Mortier} A.~M.~J.,  {Wagg} J.,
  {Dunlop} J.~S.,  {Chapin} E.~L.,  {Eales} S.~A.,  {Gazta{\~n}aga} E.,
  {Halpern} M.,  {Ivison} R.~J.,  {van Kampen} E.,  {Scott} D.,  {Serjeant} S.,
     {Smail} I.,  2007, \mnras, 379, 1571

\bibitem[\protect\citeauthoryear{{Bartelmann}, {Steinmetz} \&
  {Weiss}}{{Bartelmann} et~al.}{1995}]{B98II}
{Bartelmann} M.,  {Steinmetz} M.,    {Weiss} A.,  1995, A\&A, 297, 1

\bibitem[\protect\citeauthoryear{{Bartelmann} \& {Weiss}}{{Bartelmann} \&
  {Weiss}}{1994}]{B98I}
{Bartelmann} M.,  {Weiss} A.,  1994, A\&A, 287, 1

\bibitem[\protect\citeauthoryear{{Blain} \& {Longair}}{{Blain} \&
  {Longair}}{1993}]{Blain93}
{Blain} A.~W.,  {Longair} M.~S.,  1993, \mnras, 264, 509

\bibitem[\protect\citeauthoryear{{Blain}, {Smail}, {Ivison}, {Kneib} \&
  {Frayer}}{{Blain} et~al.}{2002}]{Blain02}
{Blain} A.~W.,  {Smail} I.,  {Ivison} R.~J.,  {Kneib} J.-P.,    {Frayer} D.~T.,
   2002, Phys.~Rep., 369, 111

\bibitem[\protect\citeauthoryear{{Borys}, {Blain}, {Dey}, {Le Floc'h},
  {Jannuzi}, {Barnard}, {Bian}, {Brodwin}, {Men{\'e}ndez-Delmestre},
  {Thompson}, {Brand}, {Brown}, {Dowell}, {Eisenhardt} \& {Farrah}}{{Borys}
  et~al.}{2006}]{Borys06}
{Borys} C.,  {Blain} A.~W.,  {Dey} A.,  {Le Floc'h} E.,  {Jannuzi} B.~T.,
  {Barnard} V.,  {Bian} C.,  {Brodwin} M.,  {Men{\'e}ndez-Delmestre} K.,
  {Thompson} D.,  {Brand} K.,  {Brown} M.~J.~I.,  {Dowell} C.~D.,  {Eisenhardt}
  P.,    {Farrah} D.,  2006, \apj, 636, 134

\bibitem[\protect\citeauthoryear{{Borys}, {Chapman}, {Donahue}, {Fahlman},
  {Halpern}, {Kneib}, {Newbury}, {Scott} \& {Smith}}{{Borys}
  et~al.}{2004}]{Borys04}
{Borys} C.,  {Chapman} S.,  {Donahue} M.,  {Fahlman} G.,  {Halpern} M.,
  {Kneib} J.-P.,  {Newbury} P.,  {Scott} D.,    {Smith} G.~P.,  2004, \mnras,
  352, 759

\bibitem[\protect\citeauthoryear{{Chapin}, {Hughes} \& {Aretxaga}}{{Chapin}
  et~al.}{2009}]{Chapin09}
{Chapin} E.~L.,  {Hughes} D.~H.,    {Aretxaga} I.,  2009, \mnras, pp 76--+

\bibitem[\protect\citeauthoryear{{Chapman}, {Blain}, {Ivison} \&
  {Smail}}{{Chapman} et~al.}{2003}]{Chapman03}
{Chapman} S.~C.,  {Blain} A.~W.,  {Ivison} R.~J.,    {Smail} I.~R.,  2003,
  \nat, 422, 695

\bibitem[\protect\citeauthoryear{{Chapman}, {Blain}, {Smail} \&
  {Ivison}}{{Chapman} et~al.}{2005}]{Chapman05}
{Chapman} S.~C.,  {Blain} A.~W.,  {Smail} I.,    {Ivison} R.~J.,  2005, \apj,
  622, 772

\bibitem[\protect\citeauthoryear{{Chary} \& {Elbaz}}{{Chary} \&
  {Elbaz}}{2001}]{Chary01}
{Chary} R.,  {Elbaz} D.,  2001, \apj, 556, 562

\bibitem[\protect\citeauthoryear{{Coppin}, {Chapin}, {Mortier}, {Scott},
  {Borys}, {Dunlop}, {Halpern}, {Hughes}, {Pope}, {Scott}, {Serjeant}, {Wagg},
  {Alexander}, {Almaini} \& {Aretxaga}}{{Coppin} et~al.}{2006}]{Coppin06}
{Coppin} K.,  {Chapin} E.~L.,  {Mortier} A.~M.~J.,  {Scott} S.~E.,  {Borys} C.,
   {Dunlop} J.~S.,  {Halpern} M.,  {Hughes} D.~H.,  {Pope} A.,  {Scott} D.,
  {Serjeant} S.,  {Wagg} J.,  {Alexander} D.~M.,  {Almaini} O.,    {Aretxaga}
  I.,  2006, \mnras, 372, 1621

\bibitem[\protect\citeauthoryear{{Courbin}, {Saha} \& {Schechter}}{{Courbin}
  et~al.}{2002}]{Courbin02}
{Courbin} F.,  {Saha} P.,    {Schechter} P.~L.,  2002, in {Courbin} F.,
  {Minniti} D.,  eds, Gravitational Lensing: An Astrophysical Tool Vol.~608 of
  Lecture Notes in Physics, Berlin Springer Verlag, {Quasar Lensing}.
pp 1--54

\bibitem[\protect\citeauthoryear{{Dole}, {Gispert}, {Lagache}, {Puget},
  {Bouchet}, {Cesarsky}, {Ciliegi}, {Clements}, {Dennefeld}, {D{\'e}sert},
  {Elbaz}, {Franceschini} \& {Guiderdoni}}{{Dole} et~al.}{2001}]{dole2001}
{Dole} H.,  {Gispert} R.,  {Lagache} G.,  {Puget} J.-L.,  {Bouchet} F.~R.,
  {Cesarsky} C.,  {Ciliegi} P.,  {Clements} D.~L.,  {Dennefeld} M.,
  {D{\'e}sert} F.-X.,  {Elbaz} D.,  {Franceschini} A.,    {Guiderdoni} B.,
  2001, A\&A, 372, 364

\bibitem[\protect\citeauthoryear{{Dole}, {Le Floc'h}, {P{\'e}rez-Gonz{\'a}lez},
  {Papovich}, {Egami}, {Lagache}, {Alonso-Herrero}, {Engelbracht}, {Gordon},
  {Hines}, {Krause} \& {Misselt}}{{Dole} et~al.}{2004}]{dole2004}
{Dole} H.,  {Le Floc'h} E.,  {P{\'e}rez-Gonz{\'a}lez} P.~G.,  {Papovich} C.,
  {Egami} E.,  {Lagache} G.,  {Alonso-Herrero} A.,  {Engelbracht} C.~W.,
  {Gordon} K.~D.,  {Hines} D.~C.,  {Krause} O.,    {Misselt} K.~A.,  2004,
  \apjs, 154, 87

\bibitem[\protect\citeauthoryear{{Dunkley}, {Spergel}, {Komatsu}, {Hinshaw},
  {Larson}, {Nolta}, {Odegard}, {Page}, {Bennett}, {Gold}, {Hill}, {Jarosik},
  {Weiland}, {Halpern}, {Kogut}, {Limon}, {Meyer}, {Tucker}, {Wollack} \&
  {Wright}}{{Dunkley} et~al.}{2008}]{Dunkley08}
{Dunkley} J.,  {Spergel} D.~N.,  {Komatsu} E.,  {Hinshaw} G.,  {Larson} D.,
  {Nolta} M.~R.,  {Odegard} N.,  {Page} L.,  {Bennett} C.~L.,  {Gold} B.,
  {Hill} R.~S.,  {Jarosik} N.,  {Weiland} J.~L.,  {Halpern} M.,  {Kogut} A.,
  {Limon} M.,  {Meyer} S.~S.,  {Tucker} G.~S.,  {Wollack} E.,    {Wright}
  E.~L.,  2008, ArXiv e-prints

\bibitem[\protect\citeauthoryear{{Dunlop}, {McLure}, {Yamada}, {Kajisawa},
  {Peacock}, {Mann}, {Hughes}, {Aretxaga}, {Muxlow}, {Richards}, {Dickinson},
  {Ivison}, {Smith}, {Smail}, {Serjeant}, {Almaini} \& {Lawrence}}{{Dunlop}
  et~al.}{2004}]{Dunlop04}
{Dunlop} J.~S.,  {McLure} R.~J.,  {Yamada} T.,  {Kajisawa} M.,  {Peacock}
  J.~A.,  {Mann} R.~G.,  {Hughes} D.~H.,  {Aretxaga} I.,  {Muxlow} T.~W.~B.,
  {Richards} A.~M.~S.,  {Dickinson} M.,  {Ivison} R.~J.,  {Smith} G.~P.,
  {Smail} I.,  {Serjeant} S.,  {Almaini} O.,    {Lawrence} A.,  2004, \mnras,
  350, 769

\bibitem[\protect\citeauthoryear{{Dunne}, {Eales}, {Edmunds}, {Ivison},
  {Alexander} \& {Clements}}{{Dunne} et~al.}{2000}]{Dunne00}
{Dunne} L.,  {Eales} S.,  {Edmunds} M.,  {Ivison} R.,  {Alexander} P.,
  {Clements} D.~L.,  2000, \mnras, 315, 115

\bibitem[\protect\citeauthoryear{{Fedeli}, {Bartelmann}, {Meneghetti} \&
  {Moscardini}}{{Fedeli} et~al.}{2008}]{Fedeli08}
{Fedeli} C.,  {Bartelmann} M.,  {Meneghetti} M.,    {Moscardini} L.,  2008,
  A\&A, 486, 35

\bibitem[\protect\citeauthoryear{{Fixsen}, {Dwek}, {Mather}, {Bennett} \&
  {Shafer}}{{Fixsen} et~al.}{1998}]{fixsen1998}
{Fixsen} D.~J.,  {Dwek} E.,  {Mather} J.~C.,  {Bennett} C.~L.,    {Shafer}
  R.~A.,  1998, \apj, 508, 123

\bibitem[\protect\citeauthoryear{{Granato}, {Silva}, {Monaco}, {Panuzzo},
  {Salucci}, {De Zotti} \& {Danese}}{{Granato} et~al.}{2001}]{Granato01}
{Granato} G.~L.,  {Silva} L.,  {Monaco} P.,  {Panuzzo} P.,  {Salucci} P.,  {De
  Zotti} G.,    {Danese} L.,  2001, \mnras, 324, 757

\bibitem[\protect\citeauthoryear{{Hilbert}, {White}, {Hartlap} \&
  {Schneider}}{{Hilbert} et~al.}{2007}]{Hilbert07}
{Hilbert} S.,  {White} S.~D.~M.,  {Hartlap} J.,    {Schneider} P.,  2007, ArXiv
  Astrophysics e-prints, 382, 121

\bibitem[\protect\citeauthoryear{{Hilbert}, {White}, {Hartlap} \&
  {Schneider}}{{Hilbert} et~al.}{2008}]{Hilbert08}
{Hilbert} S.,  {White} S.~D.~M.,  {Hartlap} J.,    {Schneider} P.,  2008,
  \mnras, 386, 1845

\bibitem[\protect\citeauthoryear{{Holland}, {MacIntosh}, {Fairley}, {Kelly},
  {Montgomery}, {Gostick}, {Atad-Ettedgui}, {Ellis}, {Robson}, {Hollister} \&
  {Woodcraft}}{{Holland} et~al.}{2006}]{Holland06}
{Holland} W.,  {MacIntosh} M.,  {Fairley} A.,  {Kelly} D.,  {Montgomery} D.,
  {Gostick} D.,  {Atad-Ettedgui} E.,  {Ellis} M.,  {Robson} I.,  {Hollister}
  M.,    {Woodcraft} A.,  2006, in Millimeter and Submillimeter Detectors and
  Instrumentation for Astronomy III. Edited by Zmuidzinas, Jonas; Holland,
  Wayne S.; Withington, Stafford; Duncan, William D. Vol.~6275 of Presented at
  the Society of Photo-Optical Instrumentation Engineers (SPIE) Conference,
  {SCUBA-2: a 10,000-pixel submillimeter camera for the James Clerk Maxwell
  Telescope}

\bibitem[\protect\citeauthoryear{{Ivison}, {Smail}, {Frayer}, {Kneib} \&
  {Blain}}{{Ivison} et~al.}{2001}]{Ivison01}
{Ivison} R.~J.,  {Smail} I.,  {Frayer} D.~T.,  {Kneib} J.-P.,    {Blain} A.~W.,
   2001, \apjl, 561, L45

\bibitem[\protect\citeauthoryear{{Ivison}, {Smail}, {Le Borgne}, {Blain},
  {Kneib}, {Bezecourt}, {Kerr} \& {Davies}}{{Ivison} et~al.}{1998}]{Ivison98}
{Ivison} R.~J.,  {Smail} I.,  {Le Borgne} J.-F.,  {Blain} A.~W.,  {Kneib}
  J.-P.,  {Bezecourt} J.,  {Kerr} T.~H.,    {Davies} J.~K.,  1998, \mnras, 298,
  583

\bibitem[\protect\citeauthoryear{{Jain}, {Seljak} \& {White}}{{Jain}
  et~al.}{2000}]{Jain00}
{Jain} B.,  {Seljak} U.,    {White} S.,  2000, \apj, 530, 547

\bibitem[\protect\citeauthoryear{{Keeton}, {Kuhlen} \& {Haiman}}{{Keeton}
  et~al.}{2005}]{Keeton04}
{Keeton} C.~R.,  {Kuhlen} M.,    {Haiman} Z.,  2005, \apj, 621, 559

\bibitem[\protect\citeauthoryear{{Kneib}, {Ellis}, {Smail}, {Couch} \&
  {Sharples}}{{Kneib} et~al.}{1996}]{Kneib96}
{Kneib} J.-P.,  {Ellis} R.~S.,  {Smail} I.,  {Couch} W.~J.,    {Sharples}
  R.~M.,  1996, \apj, 471, 643

\bibitem[\protect\citeauthoryear{{Kneib}, {van der Werf}, {Kraiberg Knudsen},
  {Smail}, {Blain}, {Frayer}, {Barnard} \& {Ivison}}{{Kneib}
  et~al.}{2004}]{Kneib04}
{Kneib} J.-P.,  {van der Werf} P.~P.,  {Kraiberg Knudsen} K.,  {Smail} I.,
  {Blain} A.,  {Frayer} D.,  {Barnard} V.,    {Ivison} R.,  2004, \mnras, 349,
  1211

\bibitem[\protect\citeauthoryear{{Kochanek}}{{Kochanek}}{2006}]{Kochanek04}
{Kochanek} C.~S.,  2006, in {Meylan} G.,  {Jetzer} P.,  {North} P.,
  {Schneider} P.,  {Kochanek} C.~S.,   {Wambsganss} J.,  eds, Saas-Fee Advanced
  Course 33: Gravitational Lensing: Strong, Weak and Micro {Part 2: Strong
  gravitational lensing}.
pp 91--268

\bibitem[\protect\citeauthoryear{{Lagache}, {Dole} \& {Puget}}{{Lagache}
  et~al.}{2003}]{Lagache03}
{Lagache} G.,  {Dole} H.,    {Puget} J.-L.,  2003, \mnras, 338, 555

\bibitem[\protect\citeauthoryear{{Lagache}, {Dole}, {Puget},
  {P{\'e}rez-Gonz{\'a}lez}, {Le Floc'h}, {Rieke}, {Papovich}, {Egami},
  {Alonso-Herrero}, {Engelbracht}, {Gordon}, {Misselt} \& {Morrison}}{{Lagache}
  et~al.}{2004}]{Lagache04}
{Lagache} G.,  {Dole} H.,  {Puget} J.-L.,  {P{\'e}rez-Gonz{\'a}lez} P.~G.,  {Le
  Floc'h} E.,  {Rieke} G.~H.,  {Papovich} C.,  {Egami} E.,  {Alonso-Herrero}
  A.,  {Engelbracht} C.~W.,  {Gordon} K.~D.,  {Misselt} K.~A.,    {Morrison}
  J.~E.,  2004, \apjs, 154, 112

\bibitem[\protect\citeauthoryear{{Lewis}, {Chapman} \& {Helou}}{{Lewis}
  et~al.}{2005}]{Lewis05}
{Lewis} G.~F.,  {Chapman} S.~C.,    {Helou} G.,  2005, \apj, 621, 32

\bibitem[\protect\citeauthoryear{{Li}, {Mao}, {Jing}, {Lin} \& {Oguri}}{{Li}
  et~al.}{2007}]{Li07}
{Li} G.~L.,  {Mao} S.,  {Jing} Y.~P.,  {Lin} W.~P.,    {Oguri} M.,  2007,
  \mnras, 378, 469

\bibitem[\protect\citeauthoryear{{Navarro}, {Frenk} \& {White}}{{Navarro}
  et~al.}{1997}]{Navarro97}
{Navarro} J.~F.,  {Frenk} C.~S.,    {White} S.~D.~M.,  1997, \apj, 490, 493

\bibitem[\protect\citeauthoryear{{Negrello}, {Perrotta}, {Gonz{\'a}lez},
  {Silva}, {de Zotti}, {Granato}, {Baccigalupi} \& {Danese}}{{Negrello}
  et~al.}{2007}]{Negrello07}
{Negrello} M.,  {Perrotta} F.,  {Gonz{\'a}lez} J.~G.-N.,  {Silva} L.,  {de
  Zotti} G.,  {Granato} G.~L.,  {Baccigalupi} C.,    {Danese} L.,  2007,
  \mnras, 377, 1557

\bibitem[\protect\citeauthoryear{{Perrotta}, {Baccigalupi}, {Bartelmann}, {De
  Zotti} \& {Granato}}{{Perrotta} et~al.}{2002}]{Perrotta02}
{Perrotta} F.,  {Baccigalupi} C.,  {Bartelmann} M.,  {De Zotti} G.,
  {Granato} G.~L.,  2002, \mnras, 329, 445

\bibitem[\protect\citeauthoryear{{Perrotta}, {Magliocchetti}, {Baccigalupi},
  {Bartelmann}, {De Zotti}, {Granato}, {Silva} \& {Danese}}{{Perrotta}
  et~al.}{2003}]{Perrotta03}
{Perrotta} F.,  {Magliocchetti} M.,  {Baccigalupi} C.,  {Bartelmann} M.,  {De
  Zotti} G.,  {Granato} G.~L.,  {Silva} L.,    {Danese} L.,  2003, \mnras, 338,
  623

\bibitem[\protect\citeauthoryear{{Pope}, {Borys}, {Scott}, {Conselice},
  {Dickinson} \& {Mobasher}}{{Pope} et~al.}{2005}]{Pope05}
{Pope} A.,  {Borys} C.,  {Scott} D.,  {Conselice} C.,  {Dickinson} M.,
  {Mobasher} B.,  2005, \mnras, 358, 149

\bibitem[\protect\citeauthoryear{{Pope}, {Scott}, {Dickinson}, {Chary},
  {Morrison}, {Borys}, {Sajina}, {Alexander}, {Daddi}, {Frayer}, {MacDonald} \&
  {Stern}}{{Pope} et~al.}{2006}]{Pope06}
{Pope} A.,  {Scott} D.,  {Dickinson} M.,  {Chary} R.-R.,  {Morrison} G.,
  {Borys} C.,  {Sajina} A.,  {Alexander} D.~M.,  {Daddi} E.,  {Frayer} D.,
  {MacDonald} E.,    {Stern} D.,  2006, \mnras, 370, 1185

\bibitem[\protect\citeauthoryear{{Schneider}, {Ehlers} \& {Falco}}{{Schneider}
  et~al.}{1992}]{Schneider92}
{Schneider} P.,  {Ehlers} J.,    {Falco} E.~E.,  1992, {Gravitational Lenses}.
Springer-Verlag Berlin

\bibitem[\protect\citeauthoryear{{Smail}, {Smith} \& {Ivison}}{{Smail}
  et~al.}{2005}]{Smail05}
{Smail} I.,  {Smith} G.~P.,    {Ivison} R.~J.,  2005, \apj, 631, 121

\bibitem[\protect\citeauthoryear{{Springel}, {White}, {Jenkins}, {Frenk},
  {Yoshida}, {Gao}, {Navarro}, {Thacker}, {Croton}, {Helly}, {Peacock}, {Cole},
  {Thomas}, {Couchman}, {Evrard}, {Colberg} \& {Pearce}}{{Springel}
  et~al.}{2005}]{Springel05}
{Springel} V.,  {White} S.~D.~M.,  {Jenkins} A.,  {Frenk} C.~S.,  {Yoshida} N.,
   {Gao} L.,  {Navarro} J.,  {Thacker} R.,  {Croton} D.,  {Helly} J.,
  {Peacock} J.~A.,  {Cole} S.,  {Thomas} P.,  {Couchman} H.,  {Evrard} A.,
  {Colberg} J.,    {Pearce} F.,  2005, Nature, 435, 629

\bibitem[\protect\citeauthoryear{{Swinbank}, {Smith}, {Bower}, {Bunker},
  {Smail}, {Ellis}, {Smith}, {Kneib}, {Sullivan} \&
  {Allington-Smith}}{{Swinbank} et~al.}{2003}]{Swinbank03}
{Swinbank} A.~M.,  {Smith} J.,  {Bower} R.~G.,  {Bunker} A.,  {Smail} I.,
  {Ellis} R.~S.,  {Smith} G.~P.,  {Kneib} J.-P.,  {Sullivan} M.,
  {Allington-Smith} J.,  2003, \apj, 598, 162

\bibitem[\protect\citeauthoryear{{Thompson}, {Serjeant}, {Jenness}, {Scott},
  {Ashdown}, {Brunt}, {Butner}, {Chapin}, {Chrysostomou}, {Clark}, {Clements},
  {Collett}, {Coppin}, {Coulson} \& {Dent}}{{Thompson}
  et~al.}{2007}]{Thompson07}
{Thompson} M.~A.,  {Serjeant} S.,  {Jenness} T.,  {Scott} D.,  {Ashdown} M.,
  {Brunt} C.,  {Butner} H.,  {Chapin} E.,  {Chrysostomou} A.~C.,  {Clark}
  J.~S.,  {Clements} D.,  {Collett} J.~L.,  {Coppin} K.,  {Coulson} I.~M.,
  {Dent} W.~R.~F.,  2007, ArXiv e-prints

\bibitem[\protect\citeauthoryear{{Wall}, {Pope} \& {Scott}}{{Wall}
  et~al.}{2008}]{WPS07}
{Wall} J.~V.,  {Pope} A.,    {Scott} D.,  2008, \mnras, 383, 435

\bibitem[\protect\citeauthoryear{{Wambsganss}}{{Wambsganss}}{1992}]{W92}
{Wambsganss} J.,  1992, \apj, 386, 19

\bibitem[\protect\citeauthoryear{{Wambsganss}, {Ostriker} \&
  {Bode}}{{Wambsganss} et~al.}{2008}]{Wambsganss08}
{Wambsganss} J.,  {Ostriker} J.~P.,    {Bode} P.,  2008, \apj, 676, 753

\bibitem[\protect\citeauthoryear{{Wang}, {Holz} \& {Munshi}}{{Wang}
  et~al.}{2002}]{Wang02}
{Wang} Y.,  {Holz} D.~E.,    {Munshi} D.,  2002, \apjl, 572, L15

\bibitem[\protect\citeauthoryear{{Wu} \& {Chiueh}}{{Wu} \&
  {Chiueh}}{2006}]{Wu06}
{Wu} J.~M.,  {Chiueh} T.,  2006, \apj, 639, 695

\bibitem[\protect\citeauthoryear{{Younger}, {Fazio}, {Huang}, {Yun}, {Wilson},
  {Ashby}, {Gurwell}, {Lai}, {Peck}, {Petitpas}, {Wilner}, {Iono}, {Kohno},
  {Kawabe}, {Hughes} \& {Aretxaga}}{{Younger} et~al.}{2007}]{Younger07}
{Younger} J.~D.,  {Fazio} G.~G.,  {Huang} J.-S.,  {Yun} M.~S.,  {Wilson} G.~W.,
   {Ashby} M.~L.~N.,  {Gurwell} M.~A.,  {Lai} K.,  {Peck} A.~B.,  {Petitpas}
  G.~R.,  {Wilner} D.~J.,  {Iono} D.,  {Kohno} K.,  {Kawabe} R.,  {Hughes}
  D.~H.,    {Aretxaga} I.,  2007, ArXiv e-prints, 708

\bibitem[\protect\citeauthoryear{{Yun}, {Reddy} \& {Condon}}{{Yun}
  et~al.}{2001}]{Yun01}
{Yun} M.~S.,  {Reddy} N.~A.,    {Condon} J.~J.,  2001, \apj, 554, 803

\end{thebibliography}

\appendix
\section{Evolutionary Model}
\label{model}

The evolutionary model used to determine 850\,\micron\ counts as a
function of redshift combines the local measurement of the joint far-IR
colour-luminosity distribution, $\Phi(L,C)$ ($L$ is the
42.5--122.5\,\micron\ far-IR luminosity, and $C$ is the logarithm of the
60/100\,\micron\ flux density ratio) presented in \citet{Chapin09},
with a two-population evolutionary scenario similar in style to the
model of \citet{Lagache03,Lagache04}, consisting of lower-luminosity
`normal' galaxies, and higher-luminosity star-bursting galaxies.  The
constraints for this model are the source counts at 850\,\micron\
\citep{Coppin06}, and 170/160\,\micron\ \citep{dole2001,dole2004}, and
the redshift distributions of 850\,\micron\ sources from
\citet{Chapman05}.  In addition we verify that the integrated
contribution to the IR background in the range
850--160\,\micron\ does not exceed that measured by {\it COBE\/}
\citep{fixsen1998}. 

\begin{figure}
  \includegraphics[width=\figurewidth]{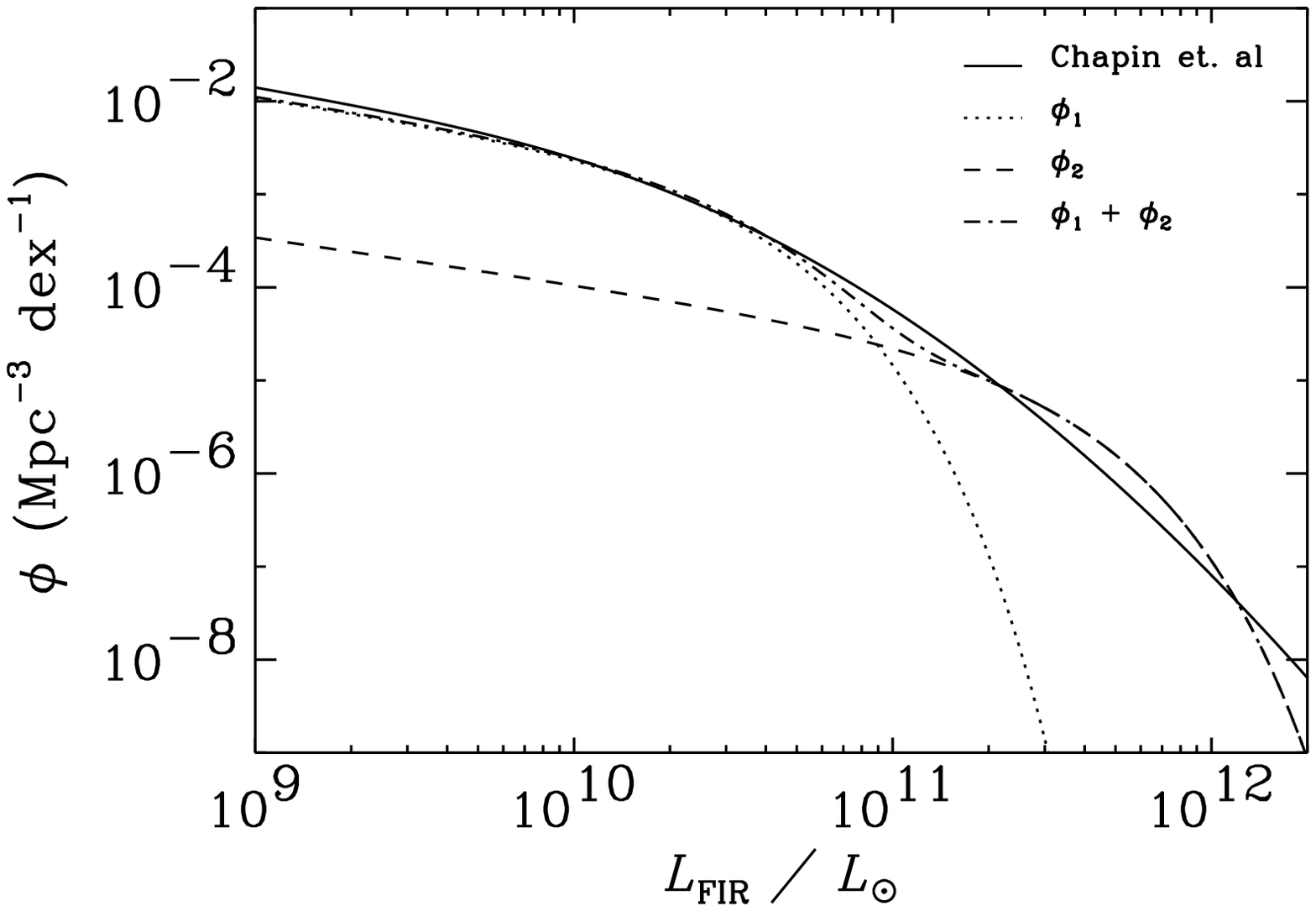}
  \caption{The local 42.5--122.5\,\micron\ far-IR luminosity function
    from \citet{Chapin09} expressed as the superposition of a
    low-luminosity `normal' galaxy population, $\phi_1(L)$, and a
    high-luminosity starburst population, $\phi_2(L)$, each
    parameterized with Schechter functions.}
  \label{fig:lf}
\end{figure}

First, similar to \citet{Yun01}, we approximate the local far-IR
luminosity function by the superposition of two Schechter functions,
$\phi(L) = \rho_*(L/L_*)^\alpha \exp(-L/L_*)$. For the
lower-luminosity objects $\rho_* = 2.3 \times
10^{-3}$\,Mpc$^{-3}$\,dex$^{-1}$ and $L_* =
2.3\times10^{10}\,{\rm L}_\odot$, while for the higher-luminosity objects
$\rho_* = 2.3 \times 10^{-5}$\,Mpc$^{-3}$\,dex$^{-1}$ and $L_* =
2.2\times10^{11}\,{\rm L}_\odot$. A value of $\alpha = -0.5$ is adopted for
both populations. The two components are compared with the original
form of \citet{Chapin09} in Figure~\ref{fig:lf}. The full local
colour-luminosity distribution for each of these populations,
$\Phi_1(L,C)$ and $\Phi_2(L,C)$ are then constructed by multiplying
the respective Schechter functions by the conditional colour
distribution $p(C|L)$, as defined in Equations~8--10 in
\citet{Chapin09} and using coefficients for the `dual power-law
luminosity function' fit.

\begin{figure}
  \includegraphics[width=\figurewidth]{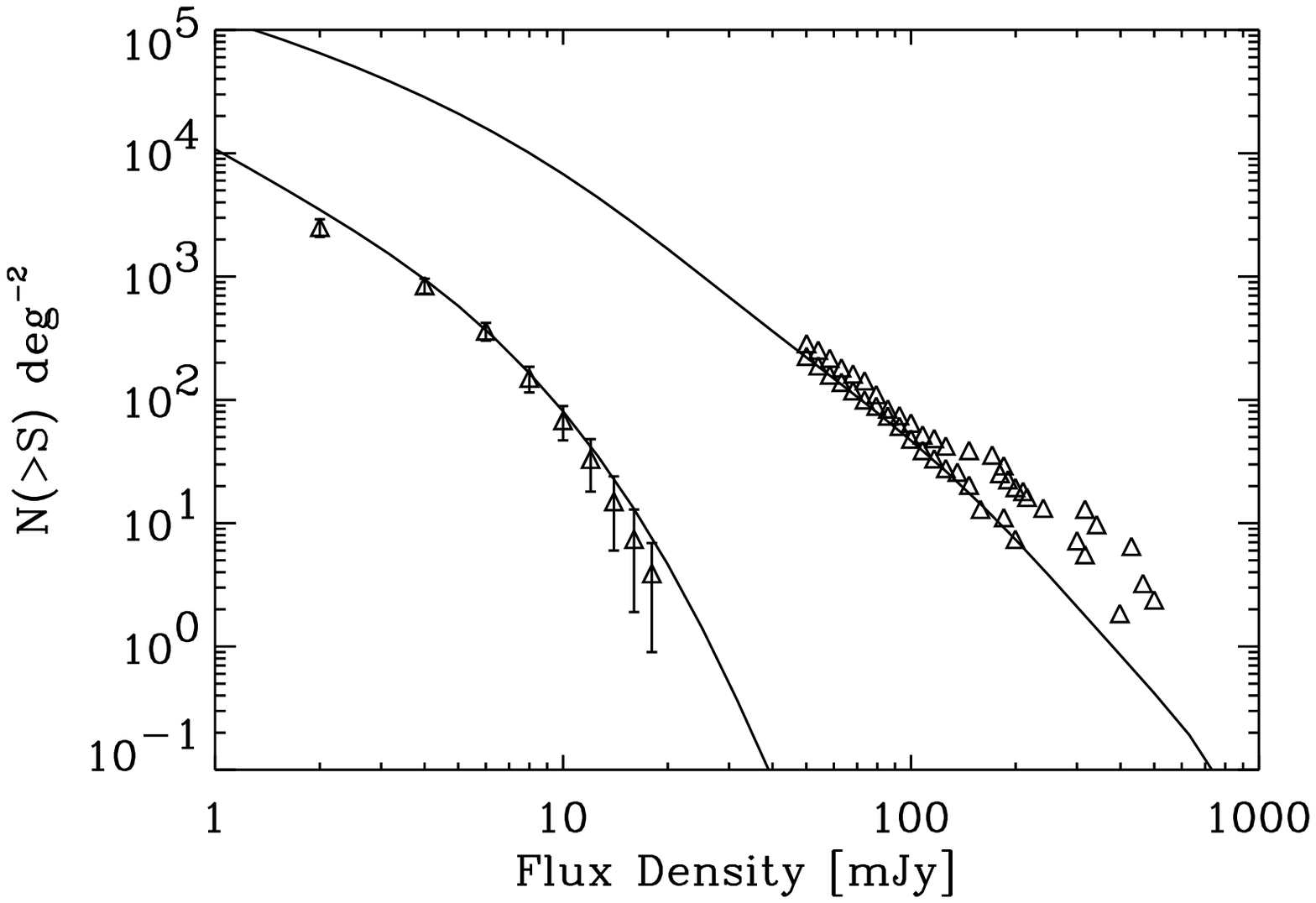}
  \caption{Observed integral source counts at 850\,\micron\
    \citep[triangles with error bars,][]{Coppin06} and at
    170/160\,\micron\ \citep[triangles without error
    bars,][]{dole2001,dole2004}. The model predictions are given by
    the solid lines.  For the purposes of this comparison the
    170\,\micron\ data are also assumed to be at 160\,\micron\ (which
    is a reasonable approximation given the proximity in wavelength).}
  \label{fig:model_counts}
\end{figure}

\begin{figure}
  \includegraphics[width=\figurewidth]{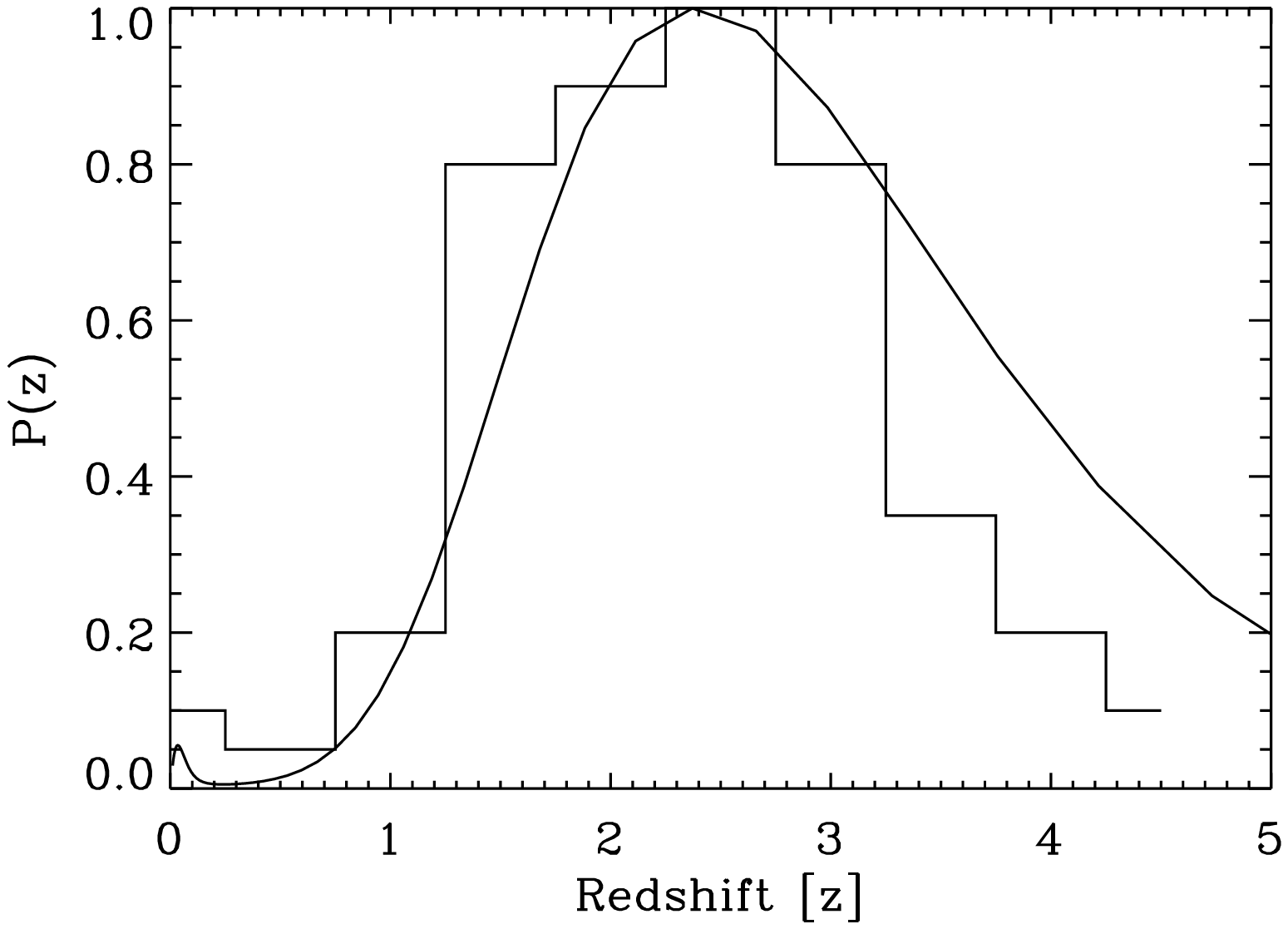}
  \caption{The observed peak-normalized redshift distribution of
    850\,\micron\ sources corrected for selection effects (histogram)
    from \citet{Chapman05}, compared with the model prediction (solid
    line) for sources with $S_{850} > 5$\,mJy. This flux limit
    approximates that of the SCUBA surveys from which the sources were
    selected for the redshift survey.}
  \label{fig:model_zdist}
\end{figure}

The `normal' galaxy population is evolved in redshift using a
combination of density and luminosity evolution, $\Phi_1(L,C,z) = g(z)
\Phi_2[L/h(z),C]$.  For the density evolution we use the form $g(z) =
(1+z)^{3/2} \mathrm{sech}^2[b\ln (1+z) - c]\cosh^2c$ from
\citet{Lewis05}, since the 2-parameter family of curves smoothly
describes a wide range of scenarios involving a steep rise with a
gradual fall.  We adopt coefficients $b=4.5$ and $c=1.6$.  For the
luminosity evolution $h(z)$ grows as $(1+z)^{2.5}$ up to $z=2.0$, remains
constant to $z=4.0$, and there are no more galaxies at higher
redshifts.  The starburst galaxy population is defined using pure
luminosity evolution, $\Phi_2(L,C,z) = \Phi_2[L/g(z),C]$, where $g(z)$
is the same parameterization used for density evolution of the
low-luminosity population, with $b=1.5$ and $c=1.3$.

Within this model the colour-luminosity correlation undergoes the same
luminosity evolution as the luminosity functions for each
population.  It was shown in \citet{Chapin09} that such evolution can
explain the observed temperatures for a sample of
850\,\micron-selected galaxies at redshifts $z>1$.

SED templates are required to calculate flux densities for particular
galaxies with luminosities $L$ and far-IR colours $C$ drawn from the
evolving distribution.  Since we only consider data in the oberved
wavelength range 850--160\,\micron\ we are free to us a simple
modified blackbody SED, $S_\nu \propto \nu^{\beta} B_\nu(T)$, as there
is no significant influence on the shape from hotter dust components,
or complicated features related to AGN activity or polycyclic aromatic
hydrocarbons in the mid-infrared. As in \citet{Chapin09} we adopt
$\beta=1.5$ such that $C$ maps directly to an effective dust
temperature $T$.

The model predictions for 850 and 160\,\micron\ source counts are
shown in Figure~\ref{fig:model_counts}, along with existing measurements,
demonstrating the agreement across this broad range in wavelengths.  In
Figure~\ref{fig:model_zdist} we compare the model redshift
distribution for 850\,\micron-selected sources brighter than 5\,mJy
compared with the redshift survey of \citet{Chapman05}, also
exhibiting good agreement.

With this model set up to be consistent with current knowledge of the local
luminosity function, submm number counts and redshift distributions, we can
use a realization of it to generate a simulated source catalogue for studying
the effects of redshift evolution on the lensing statistics.

\bsp

\end{document}